\documentclass{emulateapj}

\usepackage{amsmath}
\usepackage{amssymb}
\usepackage{graphicx}
\usepackage{color}
\usepackage{natbib}
\usepackage{epsfig}

\def\gtaprx {\lower .1ex\hbox{\rlap{\raise .6ex\hbox{\hskip .3ex
	{\ifmmode{\scriptscriptstyle >}\else
		{$\scriptscriptstyle >$}\fi}}}
	\kern -.4ex{\ifmmode{\scriptscriptstyle \sim}\else
		{$\scriptscriptstyle\sim$}\fi}}}
\def\ltaprx {\lower .1ex\hbox{\rlap{\raise .6ex\hbox{\hskip .3ex
	{\ifmmode{\scriptscriptstyle <}\else
		{$\scriptscriptstyle <$}\fi}}}
	\kern -.4ex{\ifmmode{\scriptscriptstyle \sim}\else
		{$\scriptscriptstyle\sim$}\fi}}}

\newcommand{\cutt}[1]{\textcolor{blue}{}}

\newcommand{\Ms}{{\ensuremath{{M}_{\odot} }}}

\newcommand{\HII}{{\ion{H}{2}}}

\begin{document}

\title{Finding the First Cosmic Explosions. III. Pulsational Pair-Instability Supernovae}

\author{Daniel J. Whalen\altaffilmark{1,2}, Joseph Smidt\altaffilmark{1}, Wesley 
Even\altaffilmark{3}, S. E. Woosley\altaffilmark{4}, Alexander Heger\altaffilmark{5}, 
Massimo Stiavelli\altaffilmark{6} and Chris L. Fryer\altaffilmark{3}}

\altaffiltext{1}{T-2, Los Alamos National Laboratory, Los Alamos, NM 87545, USA}

\altaffiltext{2}{Zentrum f\"{u}r Astronomie, Institut f\"{u}r Theoretische Astrophysik, 
Universit\"{a}t Heidelberg, Albert-Ueberle-Str. 2, 69120 Heidelberg, Germany}

\altaffiltext{3}{CCS-2, Los Alamos National Laboratory, Los Alamos, NM 87545, USA}

\altaffiltext{4}{Department of Astronomy and Astrophysics, UCSC, Santa Cruz, CA  
95064, USA}

\altaffiltext{5}{Monash Centre for Astrophysics, Monash University, Victoria, 3800, 
Australia}

\altaffiltext{6}{Space Telescope Science Institute, 3700 San Martin Drive, Baltimore, 
MD 21218, USA}

\begin{abstract}

Population III supernovae have been the focus of growing attention because of their 
potential to directly probe the properties of the first stars, particularly the most 
energetic events that can be seen at the edge of the observable universe.  But until 
now pulsational pair-instabilty supernovae, in which explosive thermonuclear burning 
in massive stars fails to unbind them but can eject their outer layers into space, have 
been overlooked as cosmic beacons at the earliest redshifts.  These shells can later 
collide and, like Type IIn supernovae, produce superluminous events in the UV at 
high redshifts that could be detected in the near infrared today.  We present 
numerical simulations of a 110 \Ms\ pulsational pair-instability explosion done with the 
Los Alamos radiation hydrodynamics code RAGE.  We find that collisions between 
consecutive pulsations are visible in the near infrared out to $z \sim$ 15 - 20 and can 
probe the earliest stellar populations at cosmic dawn.

\vspace{0.1in}

\end{abstract}


\section{Introduction}

Population III (Pop III) stars ended the cosmic Dark Ages and began to reionize \citep{
wan04,ket04,abs06,awb07,wa08a} and chemically enrich \citep{mbh03,ss07,bsmith09,
chiaki12,ritt12,ss13} the early intergalactic medium (IGM).  They also populated early
galaxies \citep{jgb08,get08,jlj09,get10,jeon11,pmb11,wise12,pmb12} and might be the 
origin of supermassive black holes \citep{bl03,jb07b,brmvol08,milos09,awa09,lfh09,th09,
pm11,pm12,jlj12a,wf12,agarw12,jet13,pm13,latif13c,latif13a,schl13,choi13,reis13,vol12}.  

Unfortunately, little is known for certain of the properties of the first stars. Individual Pop 
III stars will not be visible to next-generation instruments such as the \textit{James Webb 
Space Telescope} \citep[\textit{JWST};][]{jwst06} or the Thirty-Meter Telescope 
\citep[TMT; but see][about detecting the \HII\ regions of the first stars]{rz12}. Attempts to 
constrain the Pop III initial mass function (IMF) from stellar archaeology are problematic 
because of the many processes can enrich ancient, dim stars with metals over cosmic 
time \citep[e.g.,][]{Cayrel2004,bc05,fet05,Lai2008,jet09b,caffau12}.  Simulations are far 
from being able to model the formation of Pop III stars at $z \sim$ 20 from first principles 
\citep[e.g.,][]{abn02, bcl02,nu01,turk09,stacy10,clark11,hos11,sm11,get11,get12,stacy12,
hos12,susa13,hir13} and therefore do not yet predict their final masses or numbers in a 
given halo \citep[for recent reviews, see][]{dw12,glov12}.

Several groups have now turned to Pop III supernovae (SNe) as potential probes of the 
primordial IMF because they can be observed at high redshifts and their masses can be 
deduced from their light curves.  The first studies focused on pair-instability (PI) SNe, the 
extremely energetic thermonuclear explosions of 140 - 260 \Ms\ stars \citep{hw02,sc05,
wet08a,gy09,cooke12,jw11,wet13e}.  They found that Pop III PI SNe could be detected at $z 
\gtrsim$ 20 by \textit{JWST} and the TMT and at $z \sim 10 - 20$ in all-sky near infrared 
(NIR) surveys by the Wide-Field Infrared Survey Telescope (WFIRST) and the Wide-Field 
Imaging Surveyor for High-Redshift (WISH) \citep{fwf10,kasen11,wet12b,wet12a,pan12a,
hum12,det12,ds13}.  Other simulations have now shown that Pop III core-collapse (CC) 
SNe can be seen out to $z \sim$ 10 - 15 with \textit{JWST} \citep{wet12c} and that Type 
IIn SNe and supermassive thermonuclear explosions \citep{wet12d,jet13a,wet13a,wet13b} 
are visible at $z \sim$ 10 - 20.

Until now, pulsational pair-instability (PPI) SNe, in which the PI fails to unbind the star and 
instead heaves off its outer layers in a series of violent ejections, have been overlooked as 
cosmic beacons at high redshifts.  These episodes vary in energy and mass but the first 
ejection is usually the most massive one, with later pulsations ejecting less massive shells 
at higher velocities.  The later shells can overtake the first, producing collisions that are 
very luminous in the UV.   Three PPI SN candidates have been found at low redshifts, SN 
1000-1216 at $z =$ 3.9 \citep{cooke12}, SN2009ip \citep{marg13,smith13}, and perhaps 
SN 2006oz at $z =$ 0.376 \citep{lel12}.  This mechanism has also been invoked to explain 
superluminous SNe such as SN 2006gy \citep{nsmith07a,wbh07}.  At early epochs, the 
large UV fluxes of PPI SNe would be redshifted into the NIR today and could rival those of 
PI SNe.  Indeed, other types of shell-collision explosions are now known to be visible at $z 
\sim$ 10 - 20 \citep{tomin11,moriya12,tet12,wet12e,tet13}.  PPI SNe may be more 
numerous than PI SNe in the early universe, depending on the primordial IMF.

We have now modeled Pop III PPI SNe and their spectra and light curves with the Los 
Alamos RAGE and SPECTRUM codes.  In Section 2 we review the PPI SN mechanism 
and discuss our numerical methods and explosion models in Section 3.  Pair pulsations 
and their collisions are examined in Section 4.  In Section 5, we calculate NIR light curves 
for the collision in the observer frame, and we conclude in Section 6.

\section{PPI SN Explosion Mechanism}

It is generally known that Pop III stars from 140 - 260 \Ms\ die as PI SNe, but they can 
actually encounter the PI at $\sim$ 100 \Ms\ \citep[and at masses as low as 85 \Ms\ if 
they are rotating;][]{cw12}.  At these lower masses, runaway O and Si burning triggered 
by the PI may not unbind the star.  Its outer layers, which are weakly bound if it dies as 
a red supergiant, are ejected instead.  We now examine this process in greater detail 
by considering the PPI SN from \citet{wbh07} as an example.  The progenitor was a 110 
\Ms\ solar-metallicity star whose evolution was modeled in the Kepler stellar evolution 
code \citep{Weaver1978,Woosley2002}.  Mass loss was heavily suppressed over its 
lifetime to approximate the evolution of a Pop III star.  At the onset of the PI the mass of 
the star is 74.6 \Ms, with a 49.9 \Ms\ He core.  It dies as a red supergiant, with a radius 
of 1.1 $\times$ 10$^{14}$ cm and a luminosity of 9.2 $\times$ 10$^{39}$ erg s$^{-1}$.

Loss of thermal pressure support in the core due to the conversion of photons into
$e^+ - e^-$ pairs causes it to contract, radiate neutrinos and light, and grow in 
temperature from $\sim$ 10$^9$ K to 3.04 $\times$ 10$^9$ K, well above the usual 
2.0 $\times$ 10$^9$ K at which O burns stably in massive stars.  Explosive nuclear 
burning ensues, consuming 1.49 \Ms\ of O and 1.55 \Ms\ of C and releasing 1.4 
$\times$ 10$^{51}$ erg.  Most of this energy goes into expanding the star but $\sim$ 
10\% is channeled into the expulsion of the weakly bound outer layers of the core and 
surrounding envelope (24.5 \Ms, mostly He and some H).  This first shell is ejected at 
velocities of 100 - 1000 km s$^{-1}$. As shown in Fig.~2 of \citet{wbh07}, the ejection 
of the envelope looks like a weak SN, with a brief breakout luminosity of 10$^{42.7}$ 
erg s$^{-1}$ followed by a 10$^{41.9}$ erg s$^{-1}$ plateau that is powered primarily 
by He recombination.

\begin{figure*}
\begin{center}
\begin{tabular}{cc}
\epsfig{file=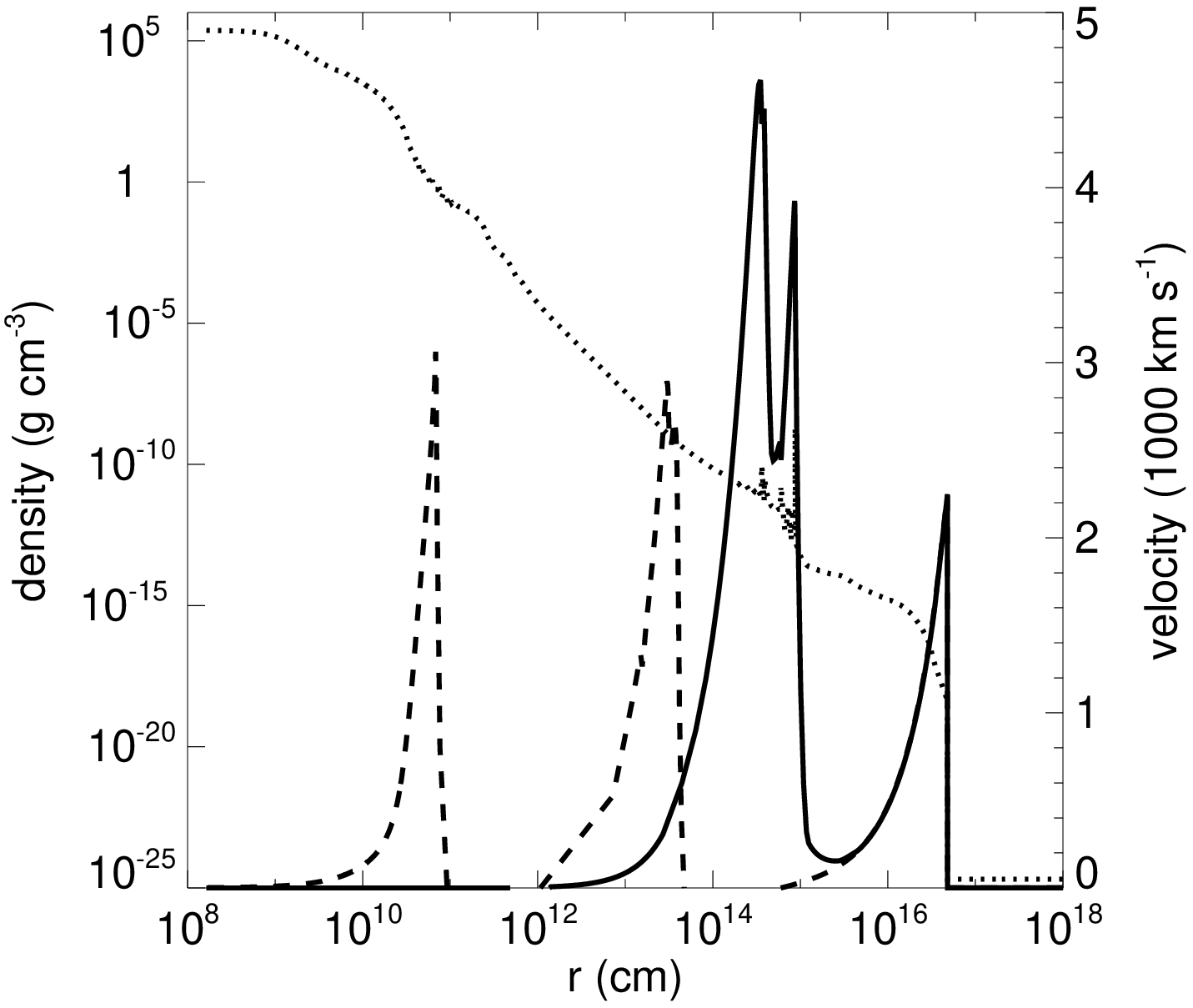,width=0.5\linewidth,clip=} & 
\epsfig{file=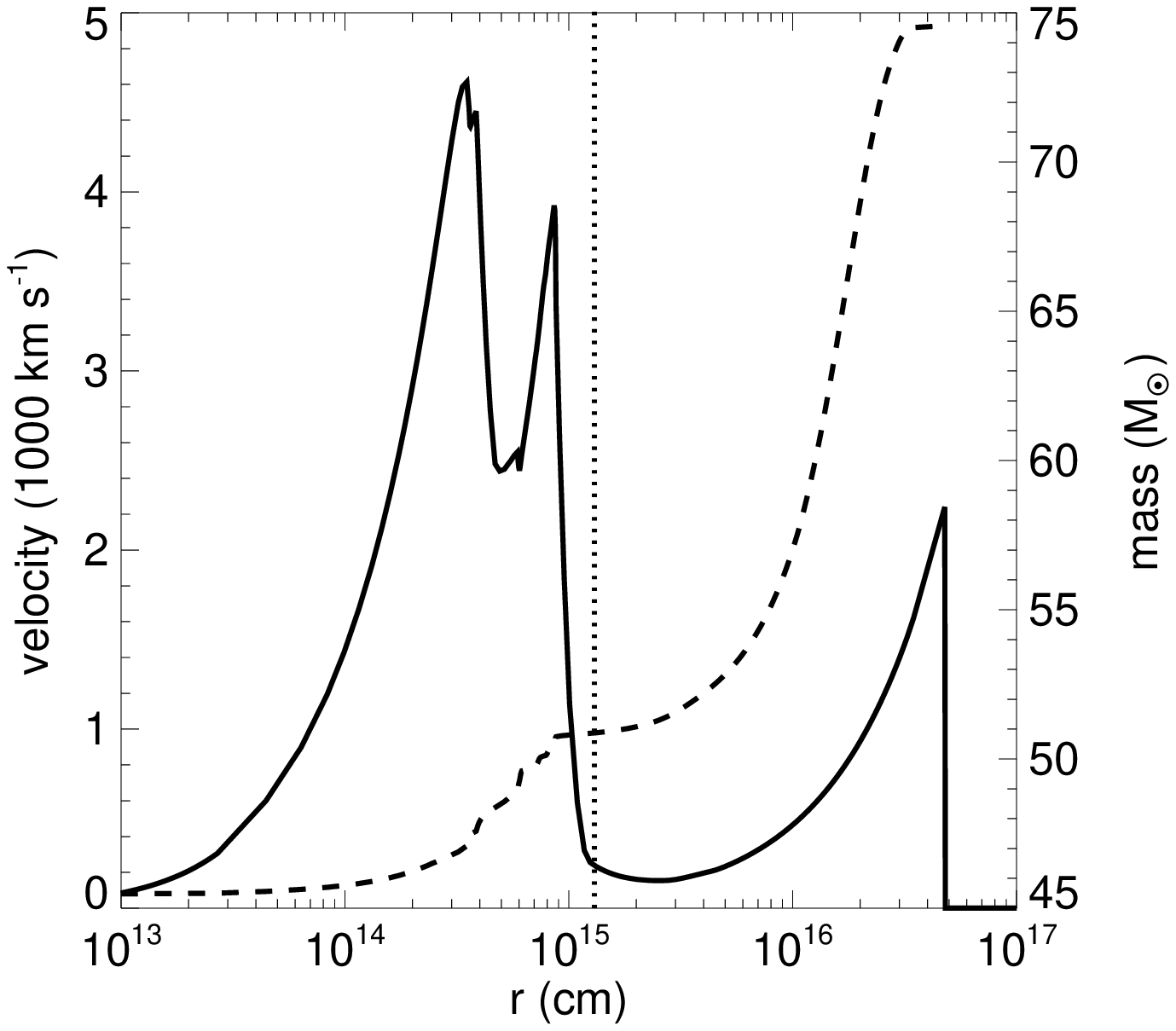,width=0.5\linewidth,clip=} \\ 
\end{tabular}
\end{center}
\caption{Evolution of the PPI.  Left:  density (dotted) and velocity (dashed) profiles for 
the star and its three pulsations just after the third ejection.  The solid line is the velocity
profile when the second and third pulses begin to overtake the first, about 120 d after
the third pulse.   Right:  enlarged view of the collision of the second and third pulsations 
with the first.  Solid line: velocities; dashed line: mass interior to the given radius; dotted:
the boundary between the second two pulses and the first.  This is the velocity profile 
that is initialized in RAGE.}
\label{fig:prof}
\end{figure*}

What remains after the initial outburst is a 50.7 \Ms\ star that is slightly more massive
than the original He core.  It again contracts, emits neutrinos and becomes hotter. 
After another 6.8 years the core again encounters the pair instability and ejects
another shell into space.  This shell is less massive, 5.1 \Ms, but more energetic,
6.0 $\times$ 10$^{50}$ erg.  It is soon followed by an even less massive but slightly 
faster shell that quickly overtakes it and collides with it, as shown in the left and right 
panels of Fig.~\ref{fig:prof}.  This initial collision is luminous but buried in an optically 
thick medium and would not be observed.  Nine years later the core contracts a final 
time, entering a stable Si burning phase that produces an Fe core that later collapses.  
If the core is rapidly rotating it could create a gamma-ray burst, either by forming a
rapidly-spinning neutron star \citep[millesecond magnetars; e.g.,][]{metz11} or a black 
hole accretion disk system \citep[collapsars; e.g.,][]{mw99}.  In this study we consider 
only the collision of the second two shells with the first and examine the NIR, radio and
x-ray signatures of Pop III GRBs elsewhere \citep{wet08b,met12a,met13}.

\section{Numerical Models}

Light curves and spectra for PPI SNe are calculated in three stages.  First, we map 
the Kepler blast profiles from \citet{wbh07} shown in the right panel of Fig.~1 into 
the RAGE code and evolve them out to 9 yr.  We then post process our RAGE 
profiles with the SPECTRUM code to construct light curves and spectra.  Finally, 
these spectra are convolved with filter response functions, cosmological redshifting 
and absorption by the neutral IGM at high $z$ to obtain NIR light curves in the 
observer frame. 

\subsection{RAGE}

We model the collision between the pair pulsations with the Los Alamos code RAGE
\citep[Radiation Adaptive Grid Eulerian;][]{rage,fet12}.  RAGE is an adaptive mesh 
refinement (AMR) radiation hydrodynamics code with a second-order conservative 
Godunov hydro scheme and grey or multigroup flux-limited diffusion for modeling 
transport in one, two, or three dimensions (1D, 2D, or 3D).  RAGE uses atomic 
opacities compiled from the Los Alamos OPLIB 
database\footnote{http://aphysics2/www.t4.lanl.gov/cgi-bin/opacity/tops.pl}\citep{
oplib} and can evolve multimaterial flows with several types of equation of state. 
Our RAGE models include multispecies advection and 2-temperature (2T) radiation
transport in which the matter and radiation temperatures, while coupled, are evolved
separately.  We also include the self-gravity of the ejected shells and the gravity due 
to the remnant star, which is treated as a point mass at the center of the coordinate 
mesh.  We evolve mass fractions for 15 elements:  H, He, C, N, O, Ne, Mg, Si, S, Ar, 
Ca, Ti, Cr, Fe and Ni.  

\subsubsection{Model Setup}

Our 1D spherical coordinate root grid has 100,000 uniform zones with an initial 
resolution of 1 $\times$ 10$^{11}$ cm and inner and outer boundaries at 2.0 $\times$ 
10$^{12}$ and 1.0 $\times$ 10$^{16}$ cm, respectively. The inner boundary is chosen 
to excise the remnant star from the grid, whose high central densities and temperatures 
would restrict the solution to unnecessarily small Courant time steps.  However, the 
gravity of this remnant, whose mass is 45.5 \Ms\ at the time the simulation is launched, 
is included in our model as a point mass at the inner boundary.  Up to 4 levels of 
refinement are applied in the initial interpolation of the profiles onto the setup grid 
and then during the simulation.  The grid is refined on the ratio of the second derivative 
to the first derivative in density, pressure and velocity according to the prescription of
\citet{Lohner1987}.

The region from the outer surface of the first pulsation to the outer boundary of the grid 
is assumed to be a diffuse ($n =$ 0.1 cm$^{-1}$) \HII\ region with a temperature of 0.01 
eV and mass fractions of 76\% H and 24\% He.  This is consistent with the general belief 
that Pop III stars ionize their halos \citep[e.g.,][]{wan04}.  We set reflecting and outflow 
boundary conditions on the fluid and radiation flows at the inner and outer boundaries of 
the mesh, respectively.  To speed up the simulation and accommodate the expansion of 
the shells we resize the grid by a factor of 2.5 every 10$^6$ time steps.  The initial time 
step on which the new series evolves scales approximately as the ratio of the new and 
old zone sizes.  We again allow up to 4 levels of refinement when mapping the flow to a 
new grid and throughout the run thereafter.    

\subsection{SPECTRUM} 

\begin{figure}
\begin{center}
\begin{tabular}{c}
\epsfig{file=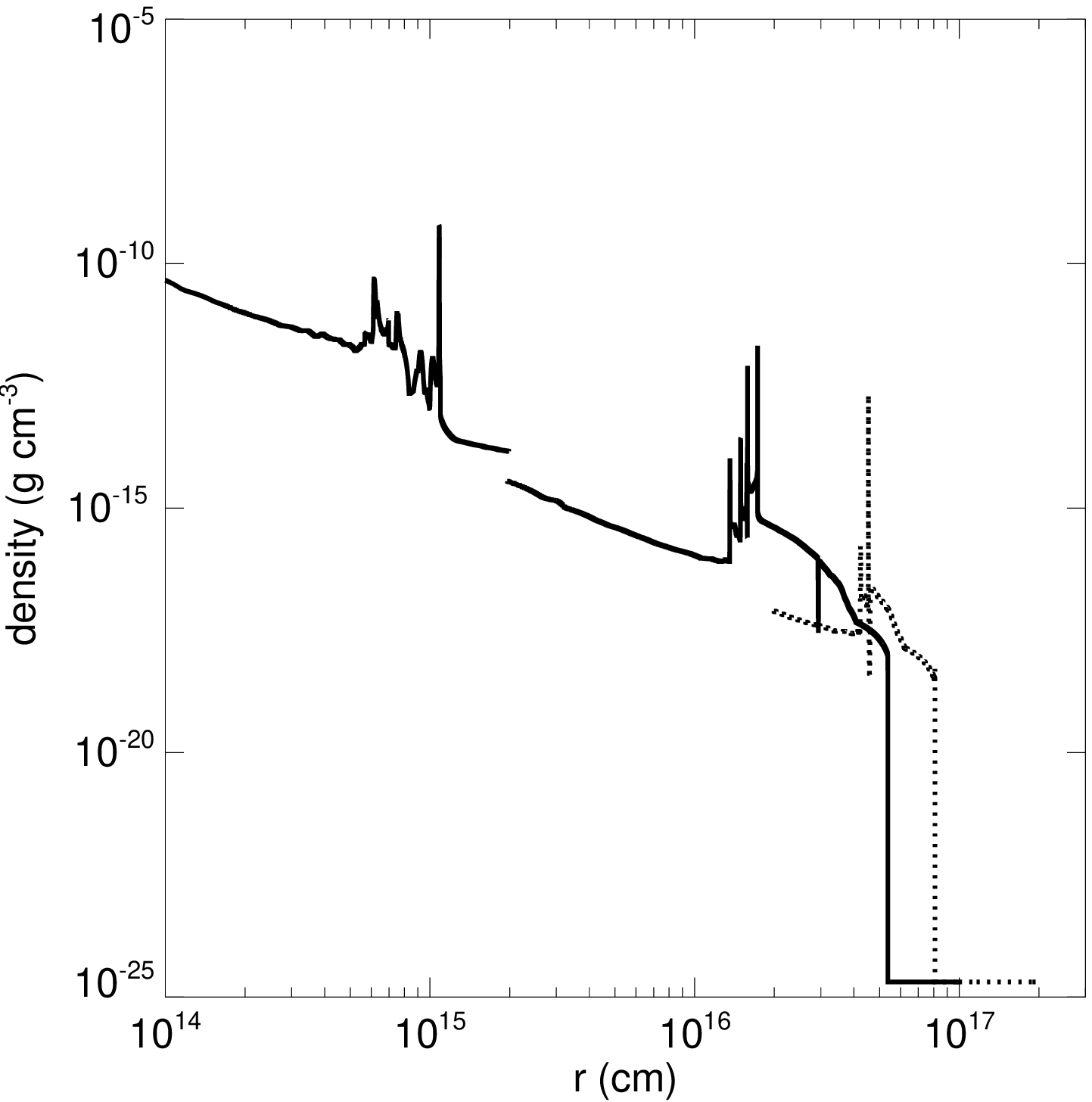,width=0.85\linewidth,clip=} \\ 
\epsfig{file=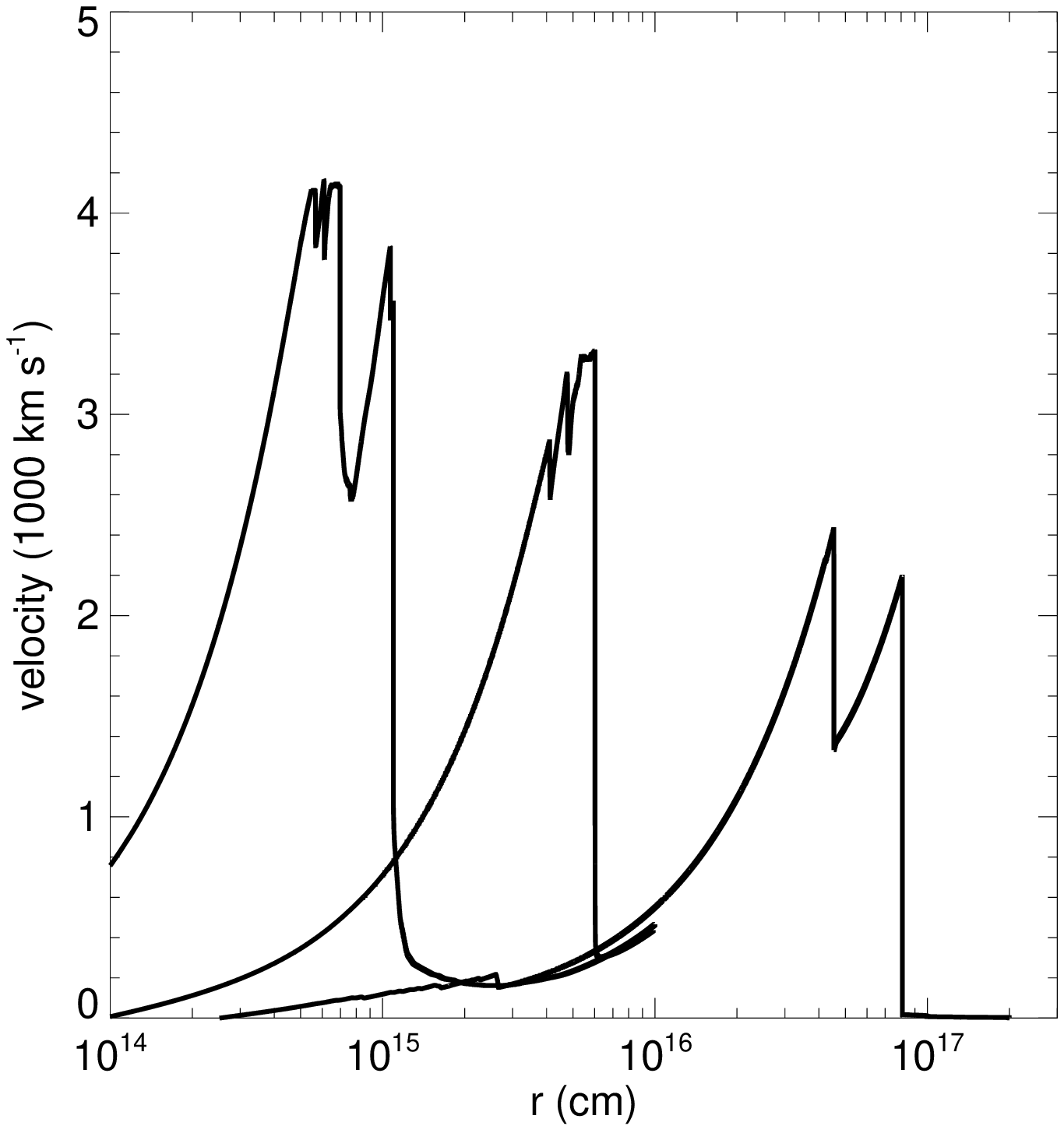,width=0.85\linewidth,clip=} \\
\epsfig{file=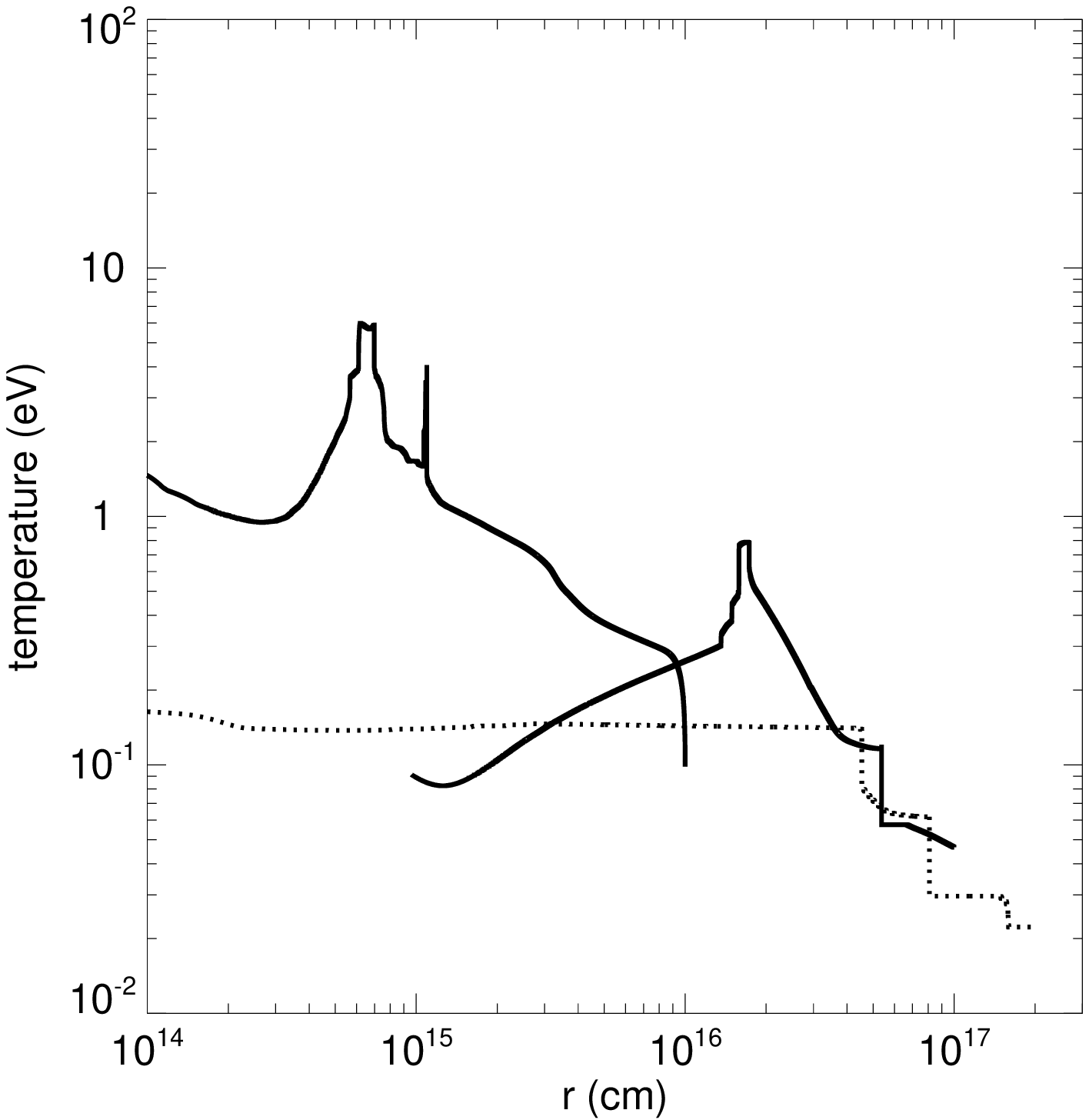,width=0.85\linewidth,clip=} 
\end{tabular}
\end{center}
\caption{The collision between the second two pair pulsations and the first.  Top:   
densities; center: velocities; bottom:  temperatures.  In the top panel, from left to right 
the times are 69 days, 1.6 yr, and 5.4 yr.  In the bottom two panels, from left to right
the times are 69 days, 5.1 months, and 5.4 yr.}
\label{fig:hydro}
\end{figure}

\begin{figure*}
\plottwo{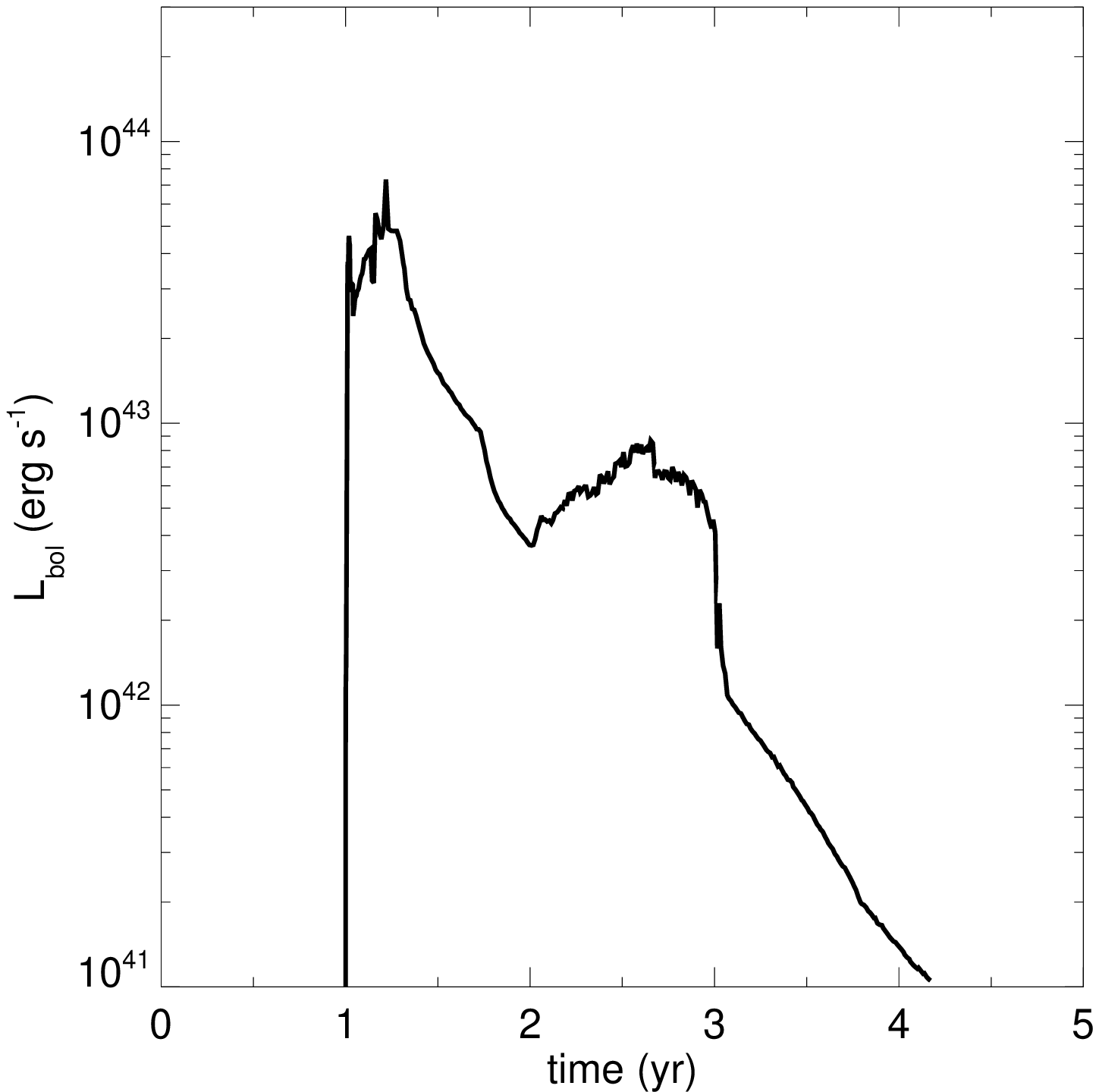}{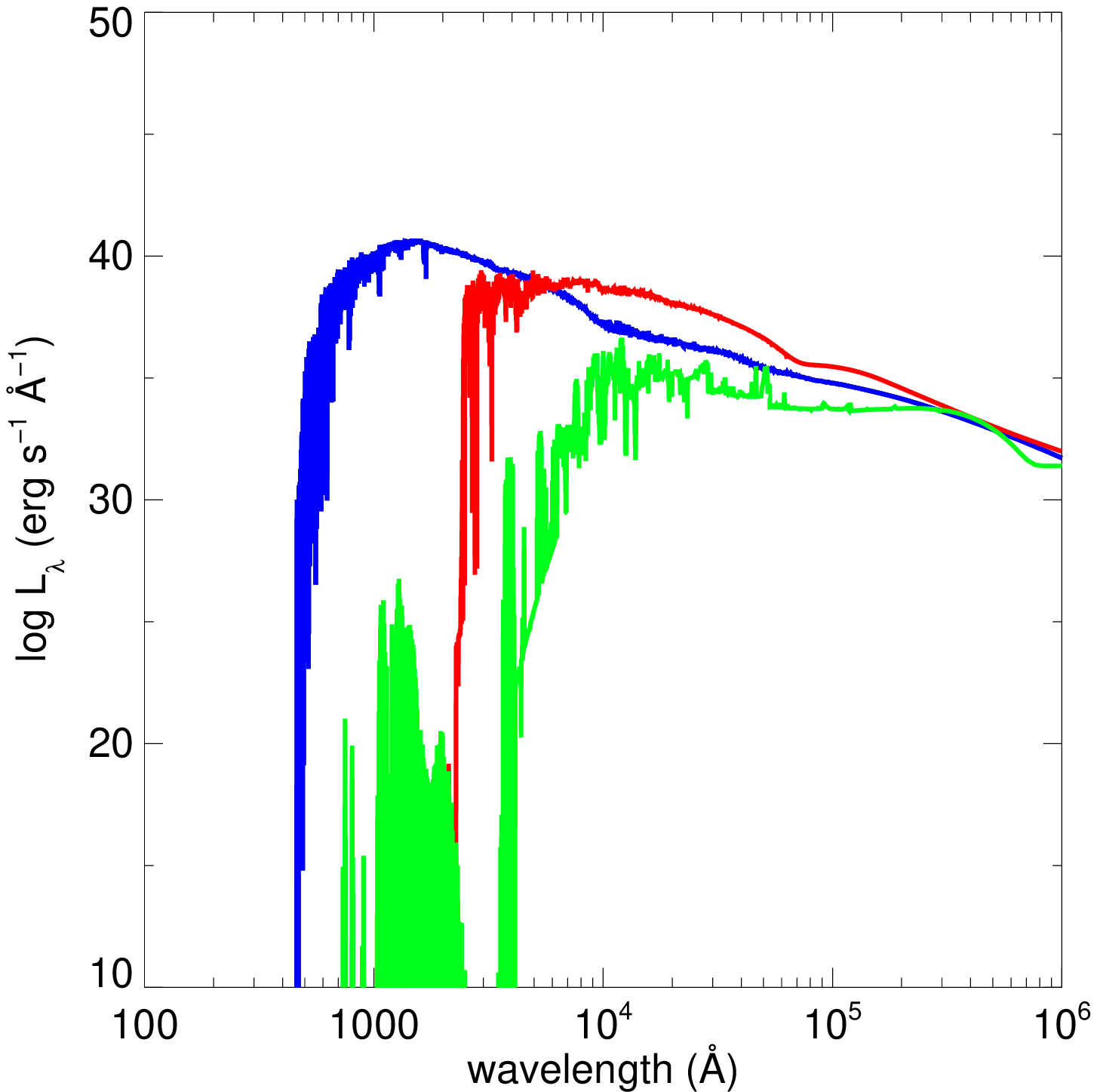}
\caption{Left panel:  bolometric luminosities for the pair pulsation collision.  Right panel: 
spectra for the collision at 69 days (blue), 1.6 yr (red), and 5.4 yr (green).} \vspace{0.1in}
\label{fig:bLC}
\end{figure*} 

To calculate a spectrum from a RAGE profile we map its densities, temperatures, mass
fractions and velocities onto a 2D grid in $r$ and $\mu =$ cos $\theta$ in the SPECTRUM 
code. SPECTRUM performs a direct sum of the luminosity of every fluid element in the 
discretized profile to compute the total flux escaping the ejecta along the line of sight 
at every wavelength.  SPECTRUM, which is described in detail in \citet{fet12}, includes 
Doppler shifts and time dilation due to the relativistic expansion of the ejecta.  It also
calculates intensities of emission lines and the attenuation of flux along the line of sight, 
capturing both limb darkening and absorption lines imprinted on the flux by intervening 
material in the ejecta and wind.

Gas densities, velocities, mass fractions and radiation temperatures are first extracted
from every level of the AMR hierarchy in RAGE and sequentially ordered by radius into 
separate files, with one variable per file.  Because of limitations on machine memory 
and time only a subset of these points are mapped onto the SPECTRUM grid. We then 
determine the position of the shock formed as the second two shells plow up the first, 
which is taken to be where the gas velocity rises above 2.3 $\times$ 10$^8$ cm s$^{-1}
$.  This velocity is chosen so that the inward sweep will not be halted by the first shell, 
whose peak velocity is 2.1 $\times$ 10$^8$ cm s$^{-1}$. Next, we find the radius of the 
$\tau = $ 40 surface by integrating the optical depth due to Thomson scattering in from 
the outer boundary, taking $\kappa_{Th}$ to be 0.288 for the material from the outer 
boundary up to the shock between the shells \citep[see Section 2.4 of][]{wet12c}.  This 
is the greatest depth from which radiation can escape from the collision between the
pulsations.  

The extracted fluid variables are then interpolated onto the SPECTRUM grid as follows.  
The inner mesh boundary is the same as for the RAGE grid and the outer boundary is 
10$^{18}$ cm.  Eight hundred uniform zones in log $r$ are assigned from the center of 
the grid to the $\tau =$ 40 surface, and the region from the $\tau =$ 40 surface to the 
shock is partitioned into 6200 uniform zones in $r$.  The region between the shock and 
the outer edge of the grid is divided into 500 uniform zones in log $r$ for a total of 7500 
radial bins.  The variables within each of these new radial bins are mass averaged so 
that the SPECTRUM grid captures very sharp features from the RAGE profile. The mesh 
is uniformly divided into 160 bins in $\mu$ from -1 to 1.  Our grid fully resolves regions of 
the collision from which photons can escape the flow and only lightly samples those from 
which they cannot.

\section{Collision Profiles}

Density, velocity and temperature profiles for the collision between the second two pair 
pulsations and the first are shown in Fig.~\ref{fig:hydro}.  It is clear from the density 
profiles in Figs.~\ref{fig:prof} and \ref{fig:hydro} that the first shell is trailed by a relatively 
dense wind-like shroud that roughly has an $r^{-2}$ profile but is basically stationary. As 
the second ejection plows through this envelope it rapidly decelerates and begins to 
shock the trailing edge of the first shell.  The shocked gas piles up in a thin, hot dense 
layer at the leading edge of the second ejection.  It is visible as the density spike at $\sim
$ 10$^{15}$ cm at 69 days, at 1.8 $\times$ 10$^{16}$ cm at 1.6 yr, and at 4.0 $\times$ 
10$^{16}$ cm at 5.4 yr.  This thin layer is the origin of all the luminosity from the collision, 
which we show in the left panel of Fig.~\ref{fig:bLC}.

Unlike Type IIn SNe, in which the collision between the ejecta and a shell are more 
abrupt, collisions between consecutive pair pulsations are more gradual because the
first ejection has no distinct inner surface.  Instead, the second ejection radiates more 
strongly as it becomes more mass loaded as it sweeps up the first shell.  This is why 
the luminosity ramps up over a period of $\sim$ 10 days before reaching a maximum
of $\sim$ 7.0 $\times$ 10$^{43}$ erg s$^{-1}$.  The collision radiates strongly for $
\sim$ 3 yr as the relative kinetic energy of the shells is converted into heat and then 
light.  Approximately 90\% of the kinetic energy is radiated away as the second shell 
advances from 10$^{15}$ cm to 2 $\times$ 10$^{16}$ cm.  The rebrightening that is
visible at $\sim$ 600 days is due to an abrupt pileup of gas in the shocked layer that
is visible as the density spike at 1.8 $\times$ 10$^{16}$ cm at 1.6 yr.  This bump in 
luminosity peaks at $\sim$ 10$^{43}$ erg s$^{-1}$ and lasts about 500 days.

The ripples in the bolometric luminosity, which are also manifest in the NIR light curves, 
are due to the classic radiative instability described by \citet{chev82} and \citet{imam84}.  
As the second two pulses plow up the first, a reverse shock forms, detaches from the 
forward shock, and backsteps into the flow in the frame of the forward shock.  But if the 
postshock gas can radiatively cool, as this gas does, the reverse shock loses pressure 
support and recedes back toward the forward shock.  As the forward shock plows up 
more material the cycle repeats.  The fluctuations in overall luminosity are due to this 
cycle, and since the bolometric luminosity varies by up to 50\% we infer that flux from 
the reverse shock is at times similar to that of the forward shock.  The STELLA light
curves from \citet{wbh07} also exhibit this radiative instability.  In a multidimensional 
simulation, the narrow region between the colliding shells in which the radiative 
instability occurs would likely be broken up by Rayleigh-Taylor (RT) instabilities, and 
this might reduce or remove altogether the fluctuations in the light curve.  These 
instabilities would also disrupt the density structures emitting most of the radiation from 
the collision, and it is unclear how this would change the luminosity.

RAGE and STELLA yield similar bolometric luminosities early in the collision, but there 
are some differences.  From 200 - 600 days the RAGE luminosities fall below 10$^{43}
$ erg s$^{-1}$ before rebrightening, while the STELLA light curve falls more gradually.  
There are also features in the peak luminosities in RAGE that are distinct from those in 
STELLA.  These discrepancies can be attributed to differences in resolution, atomic 
physics and hydrodynamic schemes, and opacities between the two codes. Differences 
between the opacities in OPLIB and in STELLA in particular could change photon 
diffusion times through the flow and the duration of the light curves.  However, the two 
codes exhibit similar bolometric luminosities overall.  The steeper decline in the RAGE 
light curve suggests that PPI SNe may be more easily detected at high $z$ than STELLA 
might predict because of the greater variability after redshifting.

It might be thought that the PPI SN would not appear to be a transient at high redshift
because its bolometric luminosity is relatively uniform for a year in the rest frame. This 
emission would last 10 - 20 yr for $z =$ 10 - 20 events in the observer frame.  But the 
spectra evolve considerably over this time, as we show in the right panel of 
Fig.~\ref{fig:bLC} and the temperatures in Fig.~\ref{fig:prof}.  At $\sim$ 70 days the 
shock is hottest, $\sim$ 5 eV or 55,000 K, and its spectrum cuts off at about 500 \AA.  
At this stage the collision radiates strongly in the UV, like Type IIn SNe. As the collision 
proceeds and more of the relative kinetic energy of the shells is dissipated, the shock 
cools and its spectrum softens.  At 1.6 yr, when the collision temporarily rebrightens, 
the shock has cooled to $\sim$ 11,000 K and its spectrum cuts off at $\sim$ 2000 \AA.  
By 5.4 yr the collision has cooled to $\sim$ 1500 K and its spectrum has evolved into 
the optical and IR.  At this point the shells have become transparent and their photons 
have escaped into the IGM, as shown by the flat material temperatures in 
Fig.~\ref{fig:hydro}.

The PPI SN is similar to the Type IIn SNe studied by \citet{wet12e} in that it is bright 
in the UV at early stages of the collision, when the shock has large radii, $\sim$ 10$
^{15}$ cm.  This radius is also similar to the inner radii of shells of the Type IIn events 
in \citet{wet12e} when the ejecta crashes into them.  With similar shock temperatures 
and radii upon collisions, the initial total luminosities for PPI SNe and Type IIne might 
be expected to be comparable, and inspection of Fig.~2 in \citet{wet12e} confirms 
this to be the case.  As the shock approaches radiation breakout from the first shell, 
one would expect low-energy photons to filter out first and later be followed by higher 
energy photons because opacities are generally greater at shorter wavelengths.  But 
the temperature of the shock rises in the early stages of the collision so both 
processes harden the spectrum as radiation fully emerges from the shell.

\begin{figure*}
\plotone{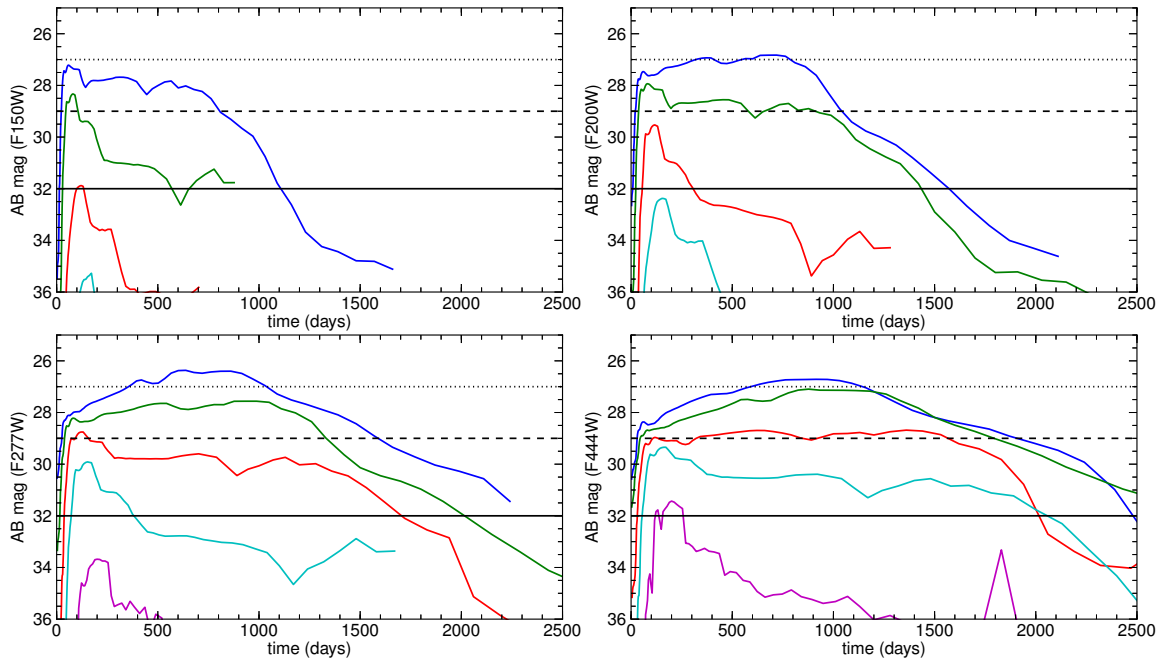}
\caption{PPI SN NIR light curves at $z =$ 7 (blue), 10 (green), 15 (red), 20 (cyan) and 
30 (purple).  The horizontal lines are photometry limits for WFIRST (dotted), WFIRST
with spectrum stacking (dashed), and \textit{JWST} (solid).} \vspace{0.1in}
\label{fig:nir}
\end{figure*} 

We note that it is primarily bound-bound and bound-free emission from the shocked 
region between the second two shells and the first that comprises the spectrum.  In
our models, the collision does not produce temperatures that are high enough for 
bremsstrahlung x-rays, as might occur in more energetic events.  The spectra are 
basically blackbody with emission and absorption lines from the shells.  These latter 
features are prominent in the spectra at later times, which appear to be blackbody 
but are sheared off at short wavelengths. 

Although our model does not produce x-rays, they might be emitted in real events by 
non local-thermodynamic-equilibrium (non-LTE) processes in the shock.  If ions in the 
shocked region between the shells decouple from electrons at early times they could 
rise to higher temperatures than in our models because they cannot efficiently radiate 
away the heat deposited by the shock.  Under these circumstances the shock might 
emit x-rays at later times.  Efforts are now underway to implement three-temperature
(3T) physics in RAGE, in which ions, electrons, and photons are evolved at separate 
temperatures, to capture non-LTE processes.  Although such x-rays would not be 
detected from high-$z$ explosions due to absorption by the neutral IGM, they might be 
visible in local events.

We also note that although the spectrum at times appears to be blackbody, one cannot
fit a blackbody to the effective temperature of the shock to estimate the NIR flux from a
high-redshift SN in lieu of an actual spectrum calculation.  The UV that is redshifted into 
the NIR comes from the short wavelength limit of the spectrum, which, unlike the longer 
wavelengths, is poorly approximated by a blackbody because of absorption by the shells.  
A blackbody fit based on the lower wavelengths would overestimate flux at these short 
wavelengths and predict too much NIR luminosity today.  

\section{NIR Light Curves / Detection Limits}

To obtain NIR light curves from the spectra, we first cosmologically redshift and dim
them.  They are then corrected for absorption by the neutral IGM at high redshift 
according to the prescription of \citet{madau95}.  Several models for absorption by
the neutral IGM exist in the literature but they yield similar transmission factors, as
discussed by \citet{ds13}.  We then convolve the spectra with filter response functions 
for {\it JWST} and WFIRST with the synthetic photometry code of \citet{su11} to 
compute NIR light curves. 

We show light curves for the PPI SN in four \textit{JWST} filters (1.5 $\mu$m,  2 $\mu$m, 
2.77 $\mu$m, and 4.44 $\mu$m at $z =$ 7, 10, 15, 20 and 30) in Fig.~\ref{fig:nir}.    The 
PPI SN is visible at $z \gtrsim$ 30 to \textit{JWST}, whose photometry limit is AB mag 32.  
The light curves all exhibit rapid rise times that correspond to the breakout of radiation 
from the first shell. Breakout is followed by a gradual rise in luminosity as the shock grows 
in radius.  But this rise and later decline has fluctuations that are due to hydrodynamics.  
As the shock expands it runs into density structures that cause it to intermittently brighten.  
Fluctuations in densities and shock temperatures also result in variations in opacities in 
front of the shock. Such features are present in Type IIn NIR light curves at early times as 
SN ejecta plows up density structures in the collision zone \citep[see Fig.~10 in][]{wet12e}.

The PPI SN shock is brighter but a little cooler than the Type IIn collisions examined
by \citet{wet12e}.  At $z =$ 15, the PPI SN is about two magnitudes brighter than the 
Type IIn, but at 4.44 $\mu$m rather than 3.56 $\mu$m.  The collision between the 
faster SN ejecta and the shell in Type IIne creates higher shock temperatures but 
happens at somewhat smaller radii, and is therefore less luminous. We note that the 
SN considered by \citet{wet12e} only had an energy of 2.4 $\times$ 10$^{51}$ erg 
and was therefore a conservative case.  In many Type IIne the ejecta can have much 
more energy and the collision could be superluminous \citep[e.g., SN 2006gy;][]{
moriya12} and nearly as bright in the NIR as a PPI SN at high redshift.

At $z \gtrsim$ 15 the PPI SN is slightly dimmer than the nominal detection limit of 
WFIRST at 4 $\mu$m, AB mag 27.  However, simple spectrum stacking could extend
this limit down to 29.  If so, WFIRST could detect these events out to $z \sim$ 20 at 2
- 4 times the rate of PI SNe, or perhaps hundreds over its mission lifetime as 
discussed in Section 6 of \citet{wet12b}.  The possibility of building up the Pop III IMF 
and tracing cosmic star formation rates at high redshift by detecting large numbers of 
primeval SNe underscores the need for all-sky NIR missions such as WFIRST and 
WISH.

We note that with only modest gravitational lensing that PPI SNe could be found in the
\textit{Cluster Lensing and Supernova Survey with \textit{Hubble}} (\textit{CLASH}) at
$z \sim$ 7 - 12 because the H band (1.63 $\mu$m) photometry limit of the Wide-Field
Camera 3 (WFC3) is $\sim$ AB mag 27.3.  However, it is not clear if there are enough
PPI SN at this epoch to be enclosed by the volume that is lensed by the clusters in the
survey.  For the mean of the star formation rates (SFRs) discussed in the next section,
\citet{wet13c} find that up to a dozen SNe from $5 < z < 12$ may already have been 
found by \textit{CLASH}.  For most IMFs it is unlikely that any of these events would be 
a PPI SN.

\section{Conclusion}

Pop III PPI SNe can be used to probe the primordial universe because they are visible 
at $z \gtrsim$ 30 to \textit{JWST} and at $z \gtrsim$ 20 to WFIRST.  They will 
complement PI SNe, Type IIne and supermassive Pop III SNe, which are visible out to 
similar redshifts.  PPI SN light curves are easily distinguished from those of other Pop III 
SNe, and their variability over likely protogalactic survey times will clearly identify them 
as transients in future surveys.  Besides revealing the properties of early stars, Pop III 
SNe will pinpoint the positions of ancient galaxies on the sky that otherwise might not 
have been discovered.  They will also constrain star formation rates in the first galaxies, 
which will yield important clues about their evolution at early times.

Although we have considered a solar-mass progenitor as a proxy for a Pop III star, this 
will not have a large effect on the energetics of the PPI or the NIR light curve of the 
collision.  The energetics of the central engine primarily depend on the entropy profile 
of the core of the star at the end of its life, which is similar for solar-metallicity and Pop
III stars of equal mass \citep[e.g.,][]{cl04,wh07} \citep[see also Fig.~1 of][]{wf12}.  But 
metals could enhance cooling in the shocked gas between the shells and lead to denser 
and thinner structures with lower temperatures and luminosities.  Metals also impose a 
more complex line structure on the spectra of the collision than would be present in 
zero-metallicity explosions.  All of this, together with the somewhat larger opacity in the 
surrounding envelope due to the metals, suggests that the NIR luminosities in our 
simulation are a lower limit to those for Pop III events.

While our fiducial model demonstrates that PPI SNe will be visible out to $z \sim$ 20 
in future observational campaigns, more work must be done to determine the range 
of stellar masses for which PPI SNe can be detected at high redshift.  Although Pop 
III stars from 85 - 140 \Ms\ can die in pulsational explosions, only those with He core 
masses from 40 - 50 \Ms\ will produce very luminous events because the timescales 
between their pulsations are short enough for the shells to actually collide (on the 
order of years).  Also, for this range in mass the collisions occur at radii of $\sim$10$
^{16}$ cm, where most of their energy is emitted in the UV and optical.  As noted 
above, RT instabilities will induce mixing at the interface between the colliding shells 
and alter their luminosity.  Future studies should be multidimensional, and focus on 
explosions in this mass range. 

Detection rates for PPI SNe at high redshift hinge on the cosmic SFR. Current estimates 
of the cosmic SFR from GRBs \citep{idf11,re12}, SNe \citep{cooke12}, early galaxies 
\citep{camp11} and simulations \citep{tfs07,ts09,wise12,jdk12,pmb12,xu13,haseg13,
mura13} vary by more than two orders of magnitude above $z \sim$ 10 \citep[see also 
Section 4 of][]{wet13c}.  Assuming fairly conservative SFRs, 5 - 10 Pop III PI SNe could 
be found by \textit{JWST} over its lifetime \citep{hum12}. Depending on the slope of the 
IMF, twice as many PPI SNe could be found.  Wide-field campaigns by WFIRST and 
WISH might discover hundreds of these events. 

\acknowledgments

We thank the anonymous referee, whose suggestions improved the quality of this paper.
DJW acknowledges support from the Baden-W\"{u}rttemberg-Stiftung by contract research 
via the programme Internationale Spitzenforschung II (grant P- LS-SPII/18).  JS and JLJ 
were supported by LANL LDRD Director's Fellowships.  MS thanks Marcia Rieke for making 
available the NIRCam filter curves and was partially supported by NASA JWST grant 
NAG5-12458.  At Santa Cruz, research was supported by NASA (NNX09AK36G) and the 
DOE-HEP Program (DE-SC0010676).  Work at LANL was done under the auspices of the 
National Nuclear Security Administration of the U.S. Department of Energy at Los Alamos 
National Laboratory under Contract No. DE-AC52-06NA25396. All RAGE and SPECTRUM 
calculations were performed on Institutional Computing (IC) and Yellow network platforms 
at LANL (Pinto, Mustang and Moonlight).

\bibliographystyle{apj}
\bibliography{refs}

\begin{thebibliography}{121}
\expandafter\ifx\csname natexlab\endcsname\relax\def\natexlab#1{#1}\fi

\bibitem[{{Abel} {et~al.}(2002){Abel}, {Bryan}, \& {Norman}}]{abn02}
{Abel}, T., {Bryan}, G.~L., \& {Norman}, M.~L. 2002, Science, 295, 93

\bibitem[{{Abel} {et~al.}(2007){Abel}, {Wise}, \& {Bryan}}]{awb07}
{Abel}, T., {Wise}, J.~H., \& {Bryan}, G.~L. 2007, \apjl, 659, L87

\bibitem[{{Agarwal} {et~al.}(2012){Agarwal}, {Khochfar}, {Johnson}, {Neistein},
  {Dalla Vecchia}, \& {Livio}}]{agarw12}
{Agarwal}, B., {Khochfar}, S., {Johnson}, J.~L., {Neistein}, E., {Dalla
  Vecchia}, C., \& {Livio}, M. 2012, \mnras, 425, 2854

\bibitem[{{Alvarez} {et~al.}(2006){Alvarez}, {Bromm}, \& {Shapiro}}]{abs06}
{Alvarez}, M.~A., {Bromm}, V., \& {Shapiro}, P.~R. 2006, \apj, 639, 621

\bibitem[{{Alvarez} {et~al.}(2009){Alvarez}, {Wise}, \& {Abel}}]{awa09}
{Alvarez}, M.~A., {Wise}, J.~H., \& {Abel}, T. 2009, \apjl, 701, L133

\bibitem[{{Beers} \& {Christlieb}(2005)}]{bc05}
{Beers}, T.~C. \& {Christlieb}, N. 2005, \araa, 43, 531

\bibitem[{{Bromm} {et~al.}(2002){Bromm}, {Coppi}, \& {Larson}}]{bcl02}
{Bromm}, V., {Coppi}, P.~S., \& {Larson}, R.~B. 2002, \apj, 564, 23

\bibitem[{{Bromm} \& {Loeb}(2003)}]{bl03}
{Bromm}, V. \& {Loeb}, A. 2003, \apj, 596, 34

\bibitem[{{Caffau} {et~al.}(2012){Caffau}, {Bonifacio}, {Fran{\c c}ois},
  {Spite}, {Spite}, {Zaggia}, {Ludwig}, {Steffen}, {Mashonkina}, {Monaco},
  {Sbordone}, {Molaro}, {Cayrel}, {Plez}, {Hill}, {Hammer}, \&
  {Randich}}]{caffau12}
{Caffau}, E., {Bonifacio}, P., {Fran{\c c}ois}, P., {Spite}, M., {Spite}, F.,
  {Zaggia}, S., {Ludwig}, H.-G., {Steffen}, M., {Mashonkina}, L., {Monaco}, L.,
  {Sbordone}, L., {Molaro}, P., {Cayrel}, R., {Plez}, B., {Hill}, V., {Hammer},
  F., \& {Randich}, S. 2012, \aap, 542, A51

\bibitem[{{Campisi} {et~al.}(2011){Campisi}, {Maio}, {Salvaterra}, \&
  {Ciardi}}]{camp11}
{Campisi}, M.~A., {Maio}, U., {Salvaterra}, R., \& {Ciardi}, B. 2011, \mnras,
  416, 2760

\bibitem[{{Cayrel} {et~al.}(2004){Cayrel}, {Depagne}, {Spite}, {Hill}, {Spite},
  {Fran{\c c}ois}, {Plez}, {Beers}, {Primas}, {Andersen}, {Barbuy},
  {Bonifacio}, {Molaro}, \& {Nordstr{\"o}m}}]{Cayrel2004}
{Cayrel}, R., {Depagne}, E., {Spite}, M., {Hill}, V., {Spite}, F., {Fran{\c
  c}ois}, P., {Plez}, B., {Beers}, T., {Primas}, F., {Andersen}, J., {Barbuy},
  B., {Bonifacio}, P., {Molaro}, P., \& {Nordstr{\"o}m}, B. 2004, \aap, 416,
  1117

\bibitem[{{Chatzopoulos} \& {Wheeler}(2012)}]{cw12}
{Chatzopoulos}, E. \& {Wheeler}, J.~C. 2012, \apj, 748, 42

\bibitem[{{Chevalier} \& {Imamura}(1982)}]{chev82}
{Chevalier}, R.~A. \& {Imamura}, J.~N. 1982, \apj, 261, 543

\bibitem[{{Chiaki} {et~al.}(2013){Chiaki}, {Yoshida}, \& {Kitayama}}]{chiaki12}
{Chiaki}, G., {Yoshida}, N., \& {Kitayama}, T. 2013, \apj, 762, 50

\bibitem[{{Chieffi} \& {Limongi}(2004)}]{cl04}
{Chieffi}, A. \& {Limongi}, M. 2004, \apj, 608, 405

\bibitem[{{Choi} {et~al.}(2013){Choi}, {Shlosman}, \& {Begelman}}]{choi13}
{Choi}, J.-H., {Shlosman}, I., \& {Begelman}, M.~C. 2013, \apj, 774, 149

\bibitem[{{Clark} {et~al.}(2011){Clark}, {Glover}, {Smith}, {Greif}, {Klessen},
  \& {Bromm}}]{clark11}
{Clark}, P.~C., {Glover}, S.~C.~O., {Smith}, R.~J., {Greif}, T.~H., {Klessen},
  R.~S., \& {Bromm}, V. 2011, Science, 331, 1040

\bibitem[{{Cooke} {et~al.}(2012){Cooke}, {Sullivan}, {Gal-Yam}, {Barton},
  {Carlberg}, {Ryan-Weber}, {Horst}, {Omori}, \& {D{\'{\i}}az}}]{cooke12}
{Cooke}, J., {Sullivan}, M., {Gal-Yam}, A., {Barton}, E.~J., {Carlberg}, R.~G.,
  {Ryan-Weber}, E.~V., {Horst}, C., {Omori}, Y., \& {D{\'{\i}}az}, C.~G. 2012,
  \nat, 491, 228

\bibitem[{{de Souza} {et~al.}(2013){de Souza}, {Ishida}, {Johnson}, {Whalen},
  \& {Mesinger}}]{ds13}
{de Souza}, R.~S., {Ishida}, E.~E.~O., {Johnson}, J.~L., {Whalen}, D.~J., \&
  {Mesinger}, A. 2013, \mnras, 436, 1555

\bibitem[{{Dessart} {et~al.}(2013){Dessart}, {Waldman}, {Livne}, {Hillier}, \&
  {Blondin}}]{det12}
{Dessart}, L., {Waldman}, R., {Livne}, E., {Hillier}, D.~J., \& {Blondin}, S.
  2013, \mnras, 428, 3227

\bibitem[{{Djorgovski} {et~al.}(2008){Djorgovski}, {Volonteri}, {Springel},
  {Bromm}, \& {Meylan}}]{brmvol08}
{Djorgovski}, S.~G., {Volonteri}, M., {Springel}, V., {Bromm}, V., \& {Meylan},
  G. 2008, in The Eleventh Marcel Grossmann Meeting On Recent Developments in
  Theoretical and Experimental General Relativity, Gravitation and Relativistic
  Field Theories, ed. {H.~Kleinert, R.~T.~Jantzen, \& R.~Ruffini}, 340--367

\bibitem[{{Frebel} {et~al.}(2005){Frebel}, {Aoki}, {Christlieb}, {Ando},
  {Asplund}, {Barklem}, {Beers}, {Eriksson}, {Fechner}, {Fujimoto}, {Honda},
  {Kajino}, {Minezaki}, {Nomoto}, {Norris}, {Ryan}, {Takada-Hidai},
  {Tsangarides}, \& {Yoshii}}]{fet05}
{Frebel}, A., {Aoki}, W., {Christlieb}, N., {Ando}, H., {Asplund}, M.,
  {Barklem}, P.~S., {Beers}, T.~C., {Eriksson}, K., {Fechner}, C., {Fujimoto},
  M.~Y., {Honda}, S., {Kajino}, T., {Minezaki}, T., {Nomoto}, K., {Norris},
  J.~E., {Ryan}, S.~G., {Takada-Hidai}, M., {Tsangarides}, S., \& {Yoshii}, Y.
  2005, \nat, 434, 871

\bibitem[{{Frey} {et~al.}(2013){Frey}, {Even}, {Whalen}, {Fryer}, {Hungerford},
  {Fontes}, \& {Colgan}}]{fet12}
{Frey}, L.~H., {Even}, W., {Whalen}, D.~J., {Fryer}, C.~L., {Hungerford},
  A.~L., {Fontes}, C.~J., \& {Colgan}, J. 2013, \apjs, 204, 16

\bibitem[{{Fryer} {et~al.}(2010){Fryer}, {Whalen}, \& {Frey}}]{fwf10}
{Fryer}, C.~L., {Whalen}, D.~J., \& {Frey}, L. 2010, in American Institute of
  Physics Conference Series, Vol. 1294, American Institute of Physics
  Conference Series, ed. D.~J. {Whalen}, V.~{Bromm}, \& N.~{Yoshida}, 70--75

\bibitem[{{Gal-Yam} {et~al.}(2009){Gal-Yam}, {Mazzali}, {Ofek}, {Nugent},
  {Kulkarni}, {Kasliwal}, {Quimby}, {Filippenko}, {Cenko}, {Chornock},
  {Waldman}, {Kasen}, {Sullivan}, {Beshore}, {Drake}, {Thomas}, {Bloom},
  {Poznanski}, {Miller}, {Foley}, {Silverman}, {Arcavi}, {Ellis}, \&
  {Deng}}]{gy09}
{Gal-Yam}, A., {Mazzali}, P., {Ofek}, E.~O., {Nugent}, P.~E., {Kulkarni},
  S.~R., {Kasliwal}, M.~M., {Quimby}, R.~M., {Filippenko}, A.~V., {Cenko},
  S.~B., {Chornock}, R., {Waldman}, R., {Kasen}, D., {Sullivan}, M., {Beshore},
  E.~C., {Drake}, A.~J., {Thomas}, R.~C., {Bloom}, J.~S., {Poznanski}, D.,
  {Miller}, A.~A., {Foley}, R.~J., {Silverman}, J.~M., {Arcavi}, I., {Ellis},
  R.~S., \& {Deng}, J. 2009, \nat, 462, 624

\bibitem[{{Gardner} {et~al.}(2006){Gardner}, {Mather}, {Clampin}, {Doyon},
  {Greenhouse}, {Hammel}, {Hutchings}, {Jakobsen}, {Lilly}, {Long}, {Lunine},
  {McCaughrean}, {Mountain}, {Nella}, {Rieke}, {Rieke}, {Rix}, {Smith},
  {Sonneborn}, {Stiavelli}, {Stockman}, {Windhorst}, \& {Wright}}]{jwst06}
{Gardner}, J.~P., {Mather}, J.~C., {Clampin}, M., {Doyon}, R., {Greenhouse},
  M.~A., {Hammel}, H.~B., {Hutchings}, J.~B., {Jakobsen}, P., {Lilly}, S.~J.,
  {Long}, K.~S., {Lunine}, J.~I., {McCaughrean}, M.~J., {Mountain}, M.,
  {Nella}, J., {Rieke}, G.~H., {Rieke}, M.~J., {Rix}, H.-W., {Smith}, E.~P.,
  {Sonneborn}, G., {Stiavelli}, M., {Stockman}, H.~S., {Windhorst}, R.~A., \&
  {Wright}, G.~S. 2006, \ssr, 123, 485

\bibitem[{{Gittings} {et~al.}(2008){Gittings}, {Weaver}, {Clover}, {Betlach},
  {Byrne}, {Coker}, {Dendy}, {Hueckstaedt}, {New}, {Oakes}, {Ranta}, \&
  {Stefan}}]{rage}
{Gittings}, M., {Weaver}, R., {Clover}, M., {Betlach}, T., {Byrne}, N.,
  {Coker}, R., {Dendy}, E., {Hueckstaedt}, R., {New}, K., {Oakes}, W.~R.,
  {Ranta}, D., \& {Stefan}, R. 2008, Computational Science and Discovery, 1,
  015005

\bibitem[{{Glover}(2013)}]{glov12}
{Glover}, S. 2013, in Astrophysics and Space Science Library, Vol. 396,
  Astrophysics and Space Science Library, ed. T.~{Wiklind}, B.~{Mobasher}, \&
  V.~{Bromm}, 103

\bibitem[{{Greif} {et~al.}(2012){Greif}, {Bromm}, {Clark}, {Glover}, {Smith},
  {Klessen}, {Yoshida}, \& {Springel}}]{get12}
{Greif}, T.~H., {Bromm}, V., {Clark}, P.~C., {Glover}, S.~C.~O., {Smith},
  R.~J., {Klessen}, R.~S., {Yoshida}, N., \& {Springel}, V. 2012, \mnras, 424,
  399

\bibitem[{{Greif} {et~al.}(2010){Greif}, {Glover}, {Bromm}, \&
  {Klessen}}]{get10}
{Greif}, T.~H., {Glover}, S.~C.~O., {Bromm}, V., \& {Klessen}, R.~S. 2010,
  \apj, 716, 510

\bibitem[{{Greif} {et~al.}(2008){Greif}, {Johnson}, {Klessen}, \&
  {Bromm}}]{get08}
{Greif}, T.~H., {Johnson}, J.~L., {Klessen}, R.~S., \& {Bromm}, V. 2008,
  \mnras, 387, 1021

\bibitem[{{Greif} {et~al.}(2011){Greif}, {Springel}, {White}, {Glover},
  {Clark}, {Smith}, {Klessen}, \& {Bromm}}]{get11}
{Greif}, T.~H., {Springel}, V., {White}, S.~D.~M., {Glover}, S.~C.~O., {Clark},
  P.~C., {Smith}, R.~J., {Klessen}, R.~S., \& {Bromm}, V. 2011, \apj, 737, 75

\bibitem[{{Hasegawa} \& {Semelin}(2013)}]{haseg13}
{Hasegawa}, K. \& {Semelin}, B. 2013, \mnras, 428, 154

\bibitem[{{Heger} \& {Woosley}(2002)}]{hw02}
{Heger}, A. \& {Woosley}, S.~E. 2002, \apj, 567, 532

\bibitem[{{Hirano} {et~al.}(2013){Hirano}, {Hosokawa}, {Yoshida}, {Umeda},
  {Omukai}, {Chiaki}, \& {Yorke}}]{hir13}
{Hirano}, S., {Hosokawa}, T., {Yoshida}, N., {Umeda}, H., {Omukai}, K.,
  {Chiaki}, G., \& {Yorke}, H.~W. 2013, arXiv:1308.4456

\bibitem[{{Hosokawa} {et~al.}(2011){Hosokawa}, {Omukai}, {Yoshida}, \&
  {Yorke}}]{hos11}
{Hosokawa}, T., {Omukai}, K., {Yoshida}, N., \& {Yorke}, H.~W. 2011, Science,
  334, 1250

\bibitem[{{Hosokawa} {et~al.}(2012){Hosokawa}, {Yoshida}, {Omukai}, \&
  {Yorke}}]{hos12}
{Hosokawa}, T., {Yoshida}, N., {Omukai}, K., \& {Yorke}, H.~W. 2012, \apjl,
  760, L37

\bibitem[{{Hummel} {et~al.}(2012){Hummel}, {Pawlik}, {Milosavljevi{\'c}}, \&
  {Bromm}}]{hum12}
{Hummel}, J.~A., {Pawlik}, A.~H., {Milosavljevi{\'c}}, M., \& {Bromm}, V. 2012,
  \apj, 755, 72

\bibitem[{{Imamura} {et~al.}(1984){Imamura}, {Wolff}, \& {Durisen}}]{imam84}
{Imamura}, J.~N., {Wolff}, M.~T., \& {Durisen}, R.~H. 1984, \apj, 276, 667

\bibitem[{{Ishida} {et~al.}(2011){Ishida}, {de Souza}, \& {Ferrara}}]{idf11}
{Ishida}, E.~E.~O., {de Souza}, R.~S., \& {Ferrara}, A. 2011, \mnras, 418, 500

\bibitem[{{Jeon} {et~al.}(2012){Jeon}, {Pawlik}, {Greif}, {Glover}, {Bromm},
  {Milosavljevi{\'c}}, \& {Klessen}}]{jeon11}
{Jeon}, M., {Pawlik}, A.~H., {Greif}, T.~H., {Glover}, S.~C.~O., {Bromm}, V.,
  {Milosavljevi{\'c}}, M., \& {Klessen}, R.~S. 2012, \apj, 754, 34

\bibitem[{{Joggerst} {et~al.}(2010){Joggerst}, {Almgren}, {Bell}, {Heger},
  {Whalen}, \& {Woosley}}]{jet09b}
{Joggerst}, C.~C., {Almgren}, A., {Bell}, J., {Heger}, A., {Whalen}, D., \&
  {Woosley}, S.~E. 2010, \apj, 709, 11

\bibitem[{{Joggerst} \& {Whalen}(2011)}]{jw11}
{Joggerst}, C.~C. \& {Whalen}, D.~J. 2011, \apj, 728, 129

\bibitem[{{Johnson} \& {Bromm}(2007)}]{jb07b}
{Johnson}, J.~L. \& {Bromm}, V. 2007, \mnras, 374, 1557

\bibitem[{{Johnson} {et~al.}(2013{\natexlab{a}}){Johnson}, {Dalla}, \&
  {Khochfar}}]{jdk12}
{Johnson}, J.~L., {Dalla}, V.~C., \& {Khochfar}, S. 2013{\natexlab{a}}, \mnras,
  428, 1857

\bibitem[{{Johnson} {et~al.}(2008){Johnson}, {Greif}, \& {Bromm}}]{jgb08}
{Johnson}, J.~L., {Greif}, T.~H., \& {Bromm}, V. 2008, \mnras, 388, 26

\bibitem[{{Johnson} {et~al.}(2009){Johnson}, {Greif}, {Bromm}, {Klessen}, \&
  {Ippolito}}]{jlj09}
{Johnson}, J.~L., {Greif}, T.~H., {Bromm}, V., {Klessen}, R.~S., \& {Ippolito},
  J. 2009, \mnras, 399, 37

\bibitem[{{Johnson} {et~al.}(2013{\natexlab{b}}){Johnson}, {Whalen}, {Even},
  {Fryer}, {Heger}, {Smidt}, \& {Chen}}]{jet13a}
{Johnson}, J.~L., {Whalen}, D.~J., {Even}, W., {Fryer}, C.~L., {Heger}, A.,
  {Smidt}, J., \& {Chen}, K.-J. 2013{\natexlab{b}}, arXiv:1304.4601

\bibitem[{{Johnson} {et~al.}(2012){Johnson}, {Whalen}, {Fryer}, \&
  {Li}}]{jlj12a}
{Johnson}, J.~L., {Whalen}, D.~J., {Fryer}, C.~L., \& {Li}, H. 2012, \apj, 750,
  66

\bibitem[{{Johnson} {et~al.}(2013{\natexlab{c}}){Johnson}, {Whalen}, {Li}, \&
  {Holz}}]{jet13}
{Johnson}, J.~L., {Whalen}, D.~J., {Li}, H., \& {Holz}, D.~E.
  2013{\natexlab{c}}, \apj, 771, 116

\bibitem[{{Kasen} {et~al.}(2011){Kasen}, {Woosley}, \& {Heger}}]{kasen11}
{Kasen}, D., {Woosley}, S.~E., \& {Heger}, A. 2011, \apj, 734, 102

\bibitem[{{Kitayama} {et~al.}(2004){Kitayama}, {Yoshida}, {Susa}, \&
  {Umemura}}]{ket04}
{Kitayama}, T., {Yoshida}, N., {Susa}, H., \& {Umemura}, M. 2004, \apj, 613,
  631

\bibitem[{{Lai} {et~al.}(2008){Lai}, {Bolte}, {Johnson}, {Lucatello}, {Heger},
  \& {Woosley}}]{Lai2008}
{Lai}, D.~K., {Bolte}, M., {Johnson}, J.~A., {Lucatello}, S., {Heger}, A., \&
  {Woosley}, S.~E. 2008, \apj, 681, 1524

\bibitem[{{Latif} {et~al.}(2013{\natexlab{a}}){Latif}, {Schleicher}, {Schmidt},
  \& {Niemeyer}}]{latif13c}
{Latif}, M.~A., {Schleicher}, D.~R.~G., {Schmidt}, W., \& {Niemeyer}, J.
  2013{\natexlab{a}}, \mnras, 433, 1607

\bibitem[{{Latif} {et~al.}(2013{\natexlab{b}}){Latif}, {Schleicher}, {Schmidt},
  \& {Niemeyer}}]{latif13a}
---. 2013{\natexlab{b}}, \mnras, 430, 588

\bibitem[{{Leloudas} {et~al.}(2012){Leloudas}, {Chatzopoulos}, {Dilday},
  {Gorosabel}, {Vinko}, {Gallazzi}, {Wheeler}, {Bassett}, {Fischer}, {Frieman},
  {Fynbo}, {Goobar}, {Jel{\'{\i}}nek}, {Malesani}, {Nichol}, {Nordin},
  {{\"O}stman}, {Sako}, {Schneider}, {Smith}, {Sollerman}, {Stritzinger},
  {Th{\"o}ne}, \& {de Ugarte Postigo}}]{lel12}
{Leloudas}, G., {Chatzopoulos}, E., {Dilday}, B., {Gorosabel}, J., {Vinko}, J.,
  {Gallazzi}, A., {Wheeler}, J.~C., {Bassett}, B., {Fischer}, J.~A., {Frieman},
  J.~A., {Fynbo}, J.~P.~U., {Goobar}, A., {Jel{\'{\i}}nek}, M., {Malesani}, D.,
  {Nichol}, R.~C., {Nordin}, J., {{\"O}stman}, L., {Sako}, M., {Schneider},
  D.~P., {Smith}, M., {Sollerman}, J., {Stritzinger}, M.~D., {Th{\"o}ne},
  C.~C., \& {de Ugarte Postigo}, A. 2012, \aap, 541, A129

\bibitem[{{Lippai} {et~al.}(2009){Lippai}, {Frei}, \& {Haiman}}]{lfh09}
{Lippai}, Z., {Frei}, Z., \& {Haiman}, Z. 2009, \apj, 701, 360

\bibitem[{L\"{o}hner(1987)}]{Lohner1987}
L\"{o}hner, R. 1987, Comput. Methods Appl. Mech. Eng., 61, 323

\bibitem[{{MacFadyen} \& {Woosley}(1999)}]{mw99}
{MacFadyen}, A.~I. \& {Woosley}, S.~E. 1999, \apj, 524, 262

\bibitem[{{Mackey} {et~al.}(2003){Mackey}, {Bromm}, \& {Hernquist}}]{mbh03}
{Mackey}, J., {Bromm}, V., \& {Hernquist}, L. 2003, \apj, 586, 1

\bibitem[{{Madau}(1995)}]{madau95}
{Madau}, P. 1995, \apj, 441, 18

\bibitem[{{Magee} {et~al.}(1995){Magee}, {Abdallah}, {Clark}, {Cohen},
  {Collins}, {Csanak}, {Fontes}, {Gauger}, {Keady}, {Kilcrease}, \&
  {Merts}}]{oplib}
{Magee}, N.~H., {Abdallah}, Jr., J., {Clark}, R.~E.~H., {Cohen}, J.~S.,
  {Collins}, L.~A., {Csanak}, G., {Fontes}, C.~J., {Gauger}, A., {Keady},
  J.~J., {Kilcrease}, D.~P., \& {Merts}, A.~L. 1995, in Astronomical Society of
  the Pacific Conference Series, Vol.~78, Astrophysical Applications of
  Powerful New Databases, ed. {S.~J.~Adelman \& W.~L.~Wiese}, 51

\bibitem[{{Margutti} {et~al.}(2013){Margutti}, {Milisavljevic}, {Soderberg},
  {Chornock}, {Zauderer}, {Murase}, {Guidorzi}, {Sanders}, {Kuin}, {Fransson},
  {Levesque}, {Chandra}, {Berger}, {Bianco}, {Brown}, {Challis},
  {Chatzopoulos}, {Cheung}, {Choi}, {Chomiuk}, {Chugai}, {Contreras}, {Drout},
  {Fesen}, {Foley}, {Fong}, {Friedman}, {Gall}, {Gehrels}, {Hjorth}, {Hsiao},
  {Kirshner}, {Im}, {Leloudas}, {Lunnan}, {Marion}, {Martin}, {Morrell},
  {Neugent}, {Omodei}, {Phillips}, {Rest}, {Silverman}, {Strader},
  {Stritzinger}, {Szalai}, {Utterback}, {Vinko}, {Wheeler}, {Arnett},
  {Campana}, {Chevalier}, {Ginsburg}, {Kamble}, {Roming}, {Pritchard}, \&
  {Stringfellow}}]{marg13}
{Margutti}, R., {Milisavljevic}, D., {Soderberg}, A.~M., {Chornock}, R.,
  {Zauderer}, B.~A., {Murase}, K., {Guidorzi}, C., {Sanders}, N.~E., {Kuin},
  P., {Fransson}, C., {Levesque}, E.~M., {Chandra}, P., {Berger}, E., {Bianco},
  F.~B., {Brown}, P.~J., {Challis}, P., {Chatzopoulos}, E., {Cheung}, C.~C.,
  {Choi}, C., {Chomiuk}, L., {Chugai}, N., {Contreras}, C., {Drout}, M.~R.,
  {Fesen}, R., {Foley}, R.~J., {Fong}, W., {Friedman}, A.~S., {Gall}, C.,
  {Gehrels}, N., {Hjorth}, J., {Hsiao}, E., {Kirshner}, R., {Im}, M.,
  {Leloudas}, G., {Lunnan}, R., {Marion}, G.~H., {Martin}, J., {Morrell}, N.,
  {Neugent}, K.~F., {Omodei}, N., {Phillips}, M.~M., {Rest}, A., {Silverman},
  J.~M., {Strader}, J., {Stritzinger}, M.~D., {Szalai}, T., {Utterback}, N.~B.,
  {Vinko}, J., {Wheeler}, J.~C., {Arnett}, D., {Campana}, S., {Chevalier}, R.,
  {Ginsburg}, A., {Kamble}, A., {Roming}, P.~W.~A., {Pritchard}, T., \&
  {Stringfellow}, G. 2013, arXiv:1306.0038

\bibitem[{{Mesler} {et~al.}(2012){Mesler}, {Whalen}, {Lloyd-Ronning}, {Fryer},
  \& {Pihlstr{\"o}m}}]{met12a}
{Mesler}, R.~A., {Whalen}, D.~J., {Lloyd-Ronning}, N.~M., {Fryer}, C.~L., \&
  {Pihlstr{\"o}m}, Y.~M. 2012, \apj, 757, 117

\bibitem[{{Mesler} {et~al.}(2013){Mesler}, {Whalen}, {Lloyd-Ronning}, {Fryer},
  \& {Pihlstr{\"o}m}}]{met13}
---. 2013, \apj, in prep

\bibitem[{{Metzger} {et~al.}(2011){Metzger}, {Giannios}, {Thompson},
  {Bucciantini}, \& {Quataert}}]{metz11}
{Metzger}, B.~D., {Giannios}, D., {Thompson}, T.~A., {Bucciantini}, N., \&
  {Quataert}, E. 2011, \mnras, 413, 2031

\bibitem[{{Milosavljevi{\'c}} {et~al.}(2009){Milosavljevi{\'c}}, {Bromm},
  {Couch}, \& {Oh}}]{milos09}
{Milosavljevi{\'c}}, M., {Bromm}, V., {Couch}, S.~M., \& {Oh}, S.~P. 2009,
  \apj, 698, 766

\bibitem[{{Moriya} {et~al.}(2013){Moriya}, {Blinnikov}, {Tominaga}, {Yoshida},
  {Tanaka}, {Maeda}, \& {Nomoto}}]{moriya12}
{Moriya}, T.~J., {Blinnikov}, S.~I., {Tominaga}, N., {Yoshida}, N., {Tanaka},
  M., {Maeda}, K., \& {Nomoto}, K. 2013, \mnras, 428, 1020

\bibitem[{{Muratov} {et~al.}(2013){Muratov}, {Gnedin}, {Gnedin}, \&
  {Zemp}}]{mura13}
{Muratov}, A.~L., {Gnedin}, O.~Y., {Gnedin}, N.~Y., \& {Zemp}, M. 2013, \apj,
  773, 19

\bibitem[{{Nakamura} \& {Umemura}(2001)}]{nu01}
{Nakamura}, F. \& {Umemura}, M. 2001, \apj, 548, 19

\bibitem[{{Pan} {et~al.}(2012){Pan}, {Kasen}, \& {Loeb}}]{pan12a}
{Pan}, T., {Kasen}, D., \& {Loeb}, A. 2012, \mnras, 422, 2701

\bibitem[{{Park} \& {Ricotti}(2011)}]{pm11}
{Park}, K. \& {Ricotti}, M. 2011, \apj, 739, 2

\bibitem[{{Park} \& {Ricotti}(2012)}]{pm12}
---. 2012, \apj, 747, 9

\bibitem[{{Park} \& {Ricotti}(2013)}]{pm13}
---. 2013, \apj, 767, 163

\bibitem[{{Pawlik} {et~al.}(2011){Pawlik}, {Milosavljevi{\'c}}, \&
  {Bromm}}]{pmb11}
{Pawlik}, A.~H., {Milosavljevi{\'c}}, M., \& {Bromm}, V. 2011, \apj, 731, 54

\bibitem[{{Pawlik} {et~al.}(2013){Pawlik}, {Milosavljevi{\'c}}, \&
  {Bromm}}]{pmb12}
---. 2013, \apj, 767, 59

\bibitem[{{Reisswig} {et~al.}(2013){Reisswig}, {Ott}, {Abdikamalov}, {Haas},
  {Moesta}, \& {Schnetter}}]{reis13}
{Reisswig}, C., {Ott}, C.~D., {Abdikamalov}, E., {Haas}, R., {Moesta}, P., \&
  {Schnetter}, E. 2013, arXiv:1304.7787

\bibitem[{{Ritter} {et~al.}(2012){Ritter}, {Safranek-Shrader}, {Gnat},
  {Milosavljevi{\'c}}, \& {Bromm}}]{ritt12}
{Ritter}, J.~S., {Safranek-Shrader}, C., {Gnat}, O., {Milosavljevi{\'c}}, M.,
  \& {Bromm}, V. 2012, \apj, 761, 56

\bibitem[{{Robertson} \& {Ellis}(2012)}]{re12}
{Robertson}, B.~E. \& {Ellis}, R.~S. 2012, \apj, 744, 95

\bibitem[{{Rydberg} {et~al.}(2013){Rydberg}, {Zackrisson}, {Lundqvist}, \&
  {Scott}}]{rz12}
{Rydberg}, C.-E., {Zackrisson}, E., {Lundqvist}, P., \& {Scott}, P. 2013,
  \mnras, 429, 3658

\bibitem[{{Safranek-Shrader} {et~al.}(2013){Safranek-Shrader}, {Milosavljevic},
  \& {Bromm}}]{ss13}
{Safranek-Shrader}, C., {Milosavljevic}, M., \& {Bromm}, V. 2013,
  arXiv:1307.1982

\bibitem[{{Scannapieco} {et~al.}(2005){Scannapieco}, {Madau}, {Woosley},
  {Heger}, \& {Ferrara}}]{sc05}
{Scannapieco}, E., {Madau}, P., {Woosley}, S., {Heger}, A., \& {Ferrara}, A.
  2005, \apj, 633, 1031

\bibitem[{{Schleicher} {et~al.}(2013){Schleicher}, {Palla}, {Ferrara}, {Galli},
  \& {Latif}}]{schl13}
{Schleicher}, D.~R.~G., {Palla}, F., {Ferrara}, A., {Galli}, D., \& {Latif}, M.
  2013, arXiv:1305.5923

\bibitem[{{Smith} \& {Sigurdsson}(2007)}]{ss07}
{Smith}, B.~D. \& {Sigurdsson}, S. 2007, \apjl, 661, L5

\bibitem[{{Smith} {et~al.}(2009){Smith}, {Turk}, {Sigurdsson}, {O'Shea}, \&
  {Norman}}]{bsmith09}
{Smith}, B.~D., {Turk}, M.~J., {Sigurdsson}, S., {O'Shea}, B.~W., \& {Norman},
  M.~L. 2009, \apj, 691, 441

\bibitem[{{Smith} {et~al.}(2013){Smith}, {Mauerhan}, \& {Prieto}}]{smith13}
{Smith}, N., {Mauerhan}, J., \& {Prieto}, J. 2013, arXiv:1308.0112

\bibitem[{{Smith} \& {McCray}(2007)}]{nsmith07a}
{Smith}, N. \& {McCray}, R. 2007, \apjl, 671, L17

\bibitem[{{Smith} {et~al.}(2011){Smith}, {Glover}, {Clark}, {Greif}, \&
  {Klessen}}]{sm11}
{Smith}, R.~J., {Glover}, S.~C.~O., {Clark}, P.~C., {Greif}, T., \& {Klessen},
  R.~S. 2011, \mnras, 414, 3633

\bibitem[{{Stacy} {et~al.}(2010){Stacy}, {Greif}, \& {Bromm}}]{stacy10}
{Stacy}, A., {Greif}, T.~H., \& {Bromm}, V. 2010, \mnras, 403, 45

\bibitem[{{Stacy} {et~al.}(2012){Stacy}, {Greif}, \& {Bromm}}]{stacy12}
---. 2012, \mnras, 422, 290

\bibitem[{{Su} {et~al.}(2011){Su}, {Stiavelli}, {Oesch}, {Trenti}, {Bergeron},
  {Bradley}, {Carollo}, {Dahlen}, {Ferguson}, {Giavalisco}, {Koekemoer},
  {Lilly}, {Lucas}, {Mobasher}, {Panagia}, \& {Pavlovsky}}]{su11}
{Su}, J., {Stiavelli}, M., {Oesch}, P., {Trenti}, M., {Bergeron}, E.,
  {Bradley}, L., {Carollo}, M., {Dahlen}, T., {Ferguson}, H.~C., {Giavalisco},
  M., {Koekemoer}, A., {Lilly}, S., {Lucas}, R.~A., {Mobasher}, B., {Panagia},
  N., \& {Pavlovsky}, C. 2011, \apj, 738, 123

\bibitem[{{Susa}(2013)}]{susa13}
{Susa}, H. 2013, \apj, 773, 185

\bibitem[{{Tanaka} {et~al.}(2013){Tanaka}, {Moriya}, \& {Yoshida}}]{tet13}
{Tanaka}, M., {Moriya}, T.~J., \& {Yoshida}, N. 2013, arXiv:1306.3743

\bibitem[{{Tanaka} {et~al.}(2012){Tanaka}, {Moriya}, {Yoshida}, \&
  {Nomoto}}]{tet12}
{Tanaka}, M., {Moriya}, T.~J., {Yoshida}, N., \& {Nomoto}, K. 2012, \mnras,
  422, 2675

\bibitem[{{Tanaka} \& {Haiman}(2009)}]{th09}
{Tanaka}, T. \& {Haiman}, Z. 2009, \apj, 696, 1798

\bibitem[{{Tominaga} {et~al.}(2011){Tominaga}, {Morokuma}, {Blinnikov},
  {Baklanov}, {Sorokina}, \& {Nomoto}}]{tomin11}
{Tominaga}, N., {Morokuma}, T., {Blinnikov}, S.~I., {Baklanov}, P., {Sorokina},
  E.~I., \& {Nomoto}, K. 2011, \apjs, 193, 20

\bibitem[{{Tornatore} {et~al.}(2007){Tornatore}, {Ferrara}, \&
  {Schneider}}]{tfs07}
{Tornatore}, L., {Ferrara}, A., \& {Schneider}, R. 2007, \mnras, 382, 945

\bibitem[{{Trenti} \& {Stiavelli}(2009)}]{ts09}
{Trenti}, M. \& {Stiavelli}, M. 2009, \apj, 694, 879

\bibitem[{{Turk} {et~al.}(2009){Turk}, {Abel}, \& {O'Shea}}]{turk09}
{Turk}, M.~J., {Abel}, T., \& {O'Shea}, B. 2009, Science, 325, 601

\bibitem[{{Volonteri}(2012)}]{vol12}
{Volonteri}, M. 2012, Science, 337, 544

\bibitem[{{Weaver} {et~al.}(1978){Weaver}, {Zimmerman}, \&
  {Woosley}}]{Weaver1978}
{Weaver}, T.~A., {Zimmerman}, G.~B., \& {Woosley}, S.~E. 1978, \apj, 225, 1021

\bibitem[{{Whalen} {et~al.}(2004){Whalen}, {Abel}, \& {Norman}}]{wan04}
{Whalen}, D., {Abel}, T., \& {Norman}, M.~L. 2004, \apj, 610, 14

\bibitem[{{Whalen} {et~al.}(2008{\natexlab{a}}){Whalen}, {O'Shea}, {Smidt}, \&
  {Norman}}]{wet08b}
{Whalen}, D., {O'Shea}, B.~W., {Smidt}, J., \& {Norman}, M.~L.
  2008{\natexlab{a}}, \apj, 679, 925

\bibitem[{{Whalen} {et~al.}(2008{\natexlab{b}}){Whalen}, {van Veelen},
  {O'Shea}, \& {Norman}}]{wet08a}
{Whalen}, D., {van Veelen}, B., {O'Shea}, B.~W., \& {Norman}, M.~L.
  2008{\natexlab{b}}, \apj, 682, 49

\bibitem[{{Whalen}(2012)}]{dw12}
{Whalen}, D.~J. 2012, arXiv:1209.4688

\bibitem[{{Whalen} {et~al.}(2013{\natexlab{a}}){Whalen}, {Even}, {Frey},
  {Smidt}, {Johnson}, {Lovekin}, {Fryer}, {Stiavelli}, {Holz}, {Heger},
  {Woosley}, \& {Hungerford}}]{wet12b}
{Whalen}, D.~J., {Even}, W., {Frey}, L.~H., {Smidt}, J., {Johnson}, J.~L.,
  {Lovekin}, C.~C., {Fryer}, C.~L., {Stiavelli}, M., {Holz}, D.~E., {Heger},
  A., {Woosley}, S.~E., \& {Hungerford}, A.~L. 2013{\natexlab{a}}, \apj, 777,
  110

\bibitem[{{Whalen} {et~al.}(2013{\natexlab{b}}){Whalen}, {Even}, {Lovekin},
  {Fryer}, {Stiavelli}, {Roming}, {Cooke}, {Pritchard}, {Holz}, \&
  {Knight}}]{wet12e}
{Whalen}, D.~J., {Even}, W., {Lovekin}, C.~C., {Fryer}, C.~L., {Stiavelli}, M.,
  {Roming}, P.~W.~A., {Cooke}, J., {Pritchard}, T.~A., {Holz}, D.~E., \&
  {Knight}, C. 2013{\natexlab{b}}, \apj, 768, 195

\bibitem[{{Whalen} {et~al.}(2013{\natexlab{c}}){Whalen}, {Even}, {Smidt},
  {Heger}, {Chen}, {Fryer}, {Stiavelli}, {Xu}, \& {Joggerst}}]{wet12d}
{Whalen}, D.~J., {Even}, W., {Smidt}, J., {Heger}, A., {Chen}, K.-J., {Fryer},
  C.~L., {Stiavelli}, M., {Xu}, H., \& {Joggerst}, C.~C. 2013{\natexlab{c}},
  \apj, 778, 17

\bibitem[{{Whalen} {et~al.}(2013{\natexlab{d}}){Whalen}, {Even}, {Smidt},
  {Heger}, {Hirschi}, {Yusof}, {Stiavelli}, {Fryer}, {Chen}, \&
  {Joggerst}}]{wet13e}
{Whalen}, D.~J., {Even}, W., {Smidt}, J., {Heger}, A., {Hirschi}, R., {Yusof},
  N., {Stiavelli}, M., {Fryer}, C.~L., {Chen}, K.-J., \& {Joggerst}, C.~C.
  2013{\natexlab{d}}, arXiv:1312.5360

\bibitem[{{Whalen} \& {Fryer}(2012)}]{wf12}
{Whalen}, D.~J. \& {Fryer}, C.~L. 2012, \apjl, 756, L19

\bibitem[{{Whalen} {et~al.}(2013{\natexlab{e}}){Whalen}, {Fryer}, {Holz},
  {Heger}, {Woosley}, {Stiavelli}, {Even}, \& {Frey}}]{wet12a}
{Whalen}, D.~J., {Fryer}, C.~L., {Holz}, D.~E., {Heger}, A., {Woosley}, S.~E.,
  {Stiavelli}, M., {Even}, W., \& {Frey}, L.~H. 2013{\natexlab{e}}, \apjl, 762,
  L6

\bibitem[{{Whalen} {et~al.}(2013{\natexlab{f}}){Whalen}, {Joggerst}, {Fryer},
  {Stiavelli}, {Heger}, \& {Holz}}]{wet12c}
{Whalen}, D.~J., {Joggerst}, C.~C., {Fryer}, C.~L., {Stiavelli}, M., {Heger},
  A., \& {Holz}, D.~E. 2013{\natexlab{f}}, \apj, 768, 95

\bibitem[{{Whalen} {et~al.}(2013{\natexlab{g}}){Whalen}, {Johnson}, {Smidt},
  {Heger}, {Even}, \& {Fryer}}]{wet13b}
{Whalen}, D.~J., {Johnson}, J.~L., {Smidt}, J., {Heger}, A., {Even}, W., \&
  {Fryer}, C.~L. 2013{\natexlab{g}}, \apj, 777, 99

\bibitem[{{Whalen} {et~al.}(2013{\natexlab{h}}){Whalen}, {Johnson}, {Smidt},
  {Meiksin}, {Heger}, {Even}, \& {Fryer}}]{wet13a}
{Whalen}, D.~J., {Johnson}, J.~L., {Smidt}, J., {Meiksin}, A., {Heger}, A.,
  {Even}, W., \& {Fryer}, C.~L. 2013{\natexlab{h}}, \apj, 774, 64

\bibitem[{{Whalen} {et~al.}(2013{\natexlab{i}}){Whalen}, {Smidt}, {Johnson},
  {Holz}, {Stiavelli}, \& {Fryer}}]{wet13c}
{Whalen}, D.~J., {Smidt}, J., {Johnson}, J.~L., {Holz}, D.~E., {Stiavelli}, M.,
  \& {Fryer}, C.~L. 2013{\natexlab{i}}, arXiv:1312.6330

\bibitem[{{Wise} \& {Abel}(2008)}]{wa08a}
{Wise}, J.~H. \& {Abel}, T. 2008, \apj, 684, 1

\bibitem[{{Wise} {et~al.}(2012){Wise}, {Turk}, {Norman}, \& {Abel}}]{wise12}
{Wise}, J.~H., {Turk}, M.~J., {Norman}, M.~L., \& {Abel}, T. 2012, \apj, 745,
  50

\bibitem[{{Woosley} {et~al.}(2007){Woosley}, {Blinnikov}, \& {Heger}}]{wbh07}
{Woosley}, S.~E., {Blinnikov}, S., \& {Heger}, A. 2007, \nat, 450, 390

\bibitem[{{Woosley} \& {Heger}(2007)}]{wh07}
{Woosley}, S.~E. \& {Heger}, A. 2007, \physrep, 442, 269

\bibitem[{{Woosley} {et~al.}(2002){Woosley}, {Heger}, \&
  {Weaver}}]{Woosley2002}
{Woosley}, S.~E., {Heger}, A., \& {Weaver}, T.~A. 2002, Reviews of Modern
  Physics, 74, 1015

\bibitem[{{Xu} {et~al.}(2013){Xu}, {Wise}, \& {Norman}}]{xu13}
{Xu}, H., {Wise}, J.~H., \& {Norman}, M.~L. 2013, \apj, 773, 83

\end{thebibliography}

\end{document}